\documentclass[usenatbib]{mnras}

\usepackage{newtxtext,newtxmath}

\usepackage[T1]{fontenc}
\usepackage{times}
\usepackage{url}
\usepackage{graphicx}	
\usepackage{graphics}	
\usepackage{epsfig}	

\usepackage{multirow}
\usepackage{xspace}
\usepackage{natbib}

\usepackage{amsmath}	
\usepackage{amssymb}	

\newcommand{\forref}[1]{{{{#1}}}}
\newcommand{\MCniter}{500\xspace}
\newcommand{\nsixte}{3\,318\xspace}
\newcommand{\nsixtefull}{14\,956\xspace}
\newcommand{\nfinal}{60\,212\xspace}
\newcommand{\noverlap}{4\,460\xspace}
\newcommand{\ndet}{4\,029\xspace} 
\newcommand{\ndetfullsky}{$\gtrsim$ 16\,000\xspace} 
\newcommand{\area}{\textcolor{black}{8\,500 and 39\,189 deg$^2$ for HECATE-SDSS and HECATE-IRAS}}
\newcommand{\nsig}{3\xspace}
	
\newcommand{\numin}{74\,182~}

\newcommand{\lx}{$L_{\rm X}$\xspace}
\newcommand{\lir}{$L_{\rm IR}$\xspace}
\newcommand{\luv}{$L_{\rm UV}$\xspace}
\newcommand{\ergs}{erg s$^{-1}$\xspace}

\newcommand{\mstar}{$M_{\star}$\xspace}
\newcommand{\msun}{$M_{\odot}$~}

\newcommand{\msunyr}{$M_{\odot}$ yr$^{-1}$\xspace}

\newcommand{\eg}{e.g.\xspace}
\newcommand{\ie}{i.e.\xspace}
\newcommand{\GALEX}{\textit{GALEX}\xspace}

\newcommand{\AllWISE}{\textit{AllWISE}\xspace}

\newcommand{\Chandra}{\textit{Chandra}\xspace}
\newcommand{\XMM}{\textit{XMM-Newton}\xspace}

\newcommand{\erosita}{\textit{eROSITA}\xspace}
\newcommand{\chandra}{\textit{Chandra}\xspace}
\newcommand{\rosat}{\textit{ROSAT}\xspace}
\newcommand{\flux}{erg s$^{-1}$ cm$^{-2}$\xspace}

\usepackage{graphicx}	
\usepackage{amsmath}	
\usepackage{amssymb}	
\usepackage{caption}

\usepackage{amsmath} 

\title{The Next Generation X-ray Galaxy Survey with \erosita}

\author[A. R. Basu-Zych et al.]{Antara R. Basu-Zych,$^{1,2}$\thanks{Email:antara.r.basu-zych@nasa.gov (ARB)}
Ann E. Hornschemeier,$^{1,3}$
Frank Haberl,$^{4}$
Neven Vulic,$^{1,5}$
\newauthor
J\"{o}rn Wilms,$^{6}$
Andreas Zezas,$^{7,8,9}$
Konstantinos Kovlakas,$^{7,9}$
Andrew Ptak$^{1,3}$
\newauthor
and Thomas Dauser$^{6}$
\\
$^{1}$NASA Goddard Space Flight Center, Code 662, Greenbelt, MD 20771, USA\\
$^{2}$Department of Physics, University of Maryland Baltimore County, Baltimore, MD 21250, USA\\
$^{3}$Department of Physics and Astronomy, Johns Hopkins University, 3400 N. Charles Street, Baltimore, MD 21218, USA\\
$^{4}$Max-Planck-Institut f{\"u}r extraterrestrische Physik, Gie{\ss}enbachstra{\ss}e 1, 85748 Garching, Germany\\
$^{5}$Department of Astronomy and Center for Space Science and Technology (CRESST), University of Maryland, College Park, MD 20742-2421, USA\\
$^{6}$Dr.\ Karl Remeis-Observatory and Erlangen Centre for Astroparticle Physics, Friedrich-Alexander-Universit\"at Erlangen-N\"urnberg, Sternwartstr.~7, 96049 \\ Bamberg, Germany\\
$^{7}$University of Crete, Physics Department \& Institute of Theoretical \& Computational Physics, 71003 Heraklion, Crete, Greece\\
$^{8}$Harvard-Smithsonian Center for Astrophysics, 60 Garden Street, Cambridge, MA 02138, USA\\
$^{9}$Institute of Astrophysics, Foundation for Research and Technology-Hellas, GR-71110 Heraklion, Greece
}

\date{Accepted 2020 July 24. Received 2020 June 29; in original form 2020 March 02}

\pubyear{2020}

\begin{document}
\label{firstpage}
\pagerange{\pageref{firstpage}--\pageref{lastpage}}
\maketitle

\begin{abstract}
\erosita, launched on 13 July 2019, will be completing the first all-sky survey in the soft and medium X-ray band in nearly three decades.  This 4-year survey, finishing in late 2023, will present a rich legacy for the entire astrophysics community and complement upcoming multi-wavelength surveys (with, \eg the Large Synoptic Survey Telescope and the Dark Energy Survey). Besides the major scientific aim to study active galactic nuclei (AGN) and galaxy clusters, \erosita will contribute significantly to X-ray studies of normal (\ie not AGN) galaxies. Starting from multi-wavelength catalogues, we measure star formation rates and stellar masses for \nfinal galaxies constrained to distances of 50--200~Mpc. We chose this distance range to focus on the relatively unexplored volume outside the local Universe, where galaxies will be largely spatially unresolved and probe a range of X-ray luminosities that overlap with the low luminosity and/or highly obscured AGN population. We use the most recent X-ray scaling relations as well as the on-orbit \erosita instrument performance to predict the X-ray emission from XRBs and diffuse hot gas and to perform both an analytic prediction and an end-to-end simulation using the mission simulation software, SIXTE. We consider potential contributions from hidden AGN and comment on the impact of normal galaxies on the measurement of the faint end of the AGN luminosity function.  We predict that the \erosita 4-year survey, will detect $\gtrsim$\forref{15\,000~}galaxies ($3\sigma$ significance) at 50--200~Mpc, which is $\sim 100\times$ more normal galaxies than detected in any X-ray survey to date. 
\end{abstract}



\section{Introduction} \label{sec:intro}
It has now been more than two decades since X-ray surveys introduced normal galaxies as a population of X-ray emitters \citep[\eg with the {\it Einstein} Observatory and the {\it ROSAT} all-sky survey,][]{Fabbiano89,Fabbiano92, Boller92,Zimmermann2001,Tajer2004}.  The X-ray emission from normal galaxies arises chiefly from accretion onto compact objects in binary systems and from the hot phase of the Interstellar Medium (ISM). 
It is important to study normal galaxies at X-ray energies for a variety of reasons.   The dominant accreting compact object population, in the absence of an accreting super-massive black hole (SMBH), are neutron star (NS) and stellar-mass black hole (BH) populations, which uniquely trace the endpoint of massive star formation.  Observations of BH and NS in accreting X-ray binaries (XRBs) tell us how they form and evolve and additionally provide understanding of the binary phase of stellar evolution.  Further, the collective X-ray output from XRBs can rival that of accreting supermassive BH (SMBH or Active Galactic Nuclei, AGN) at the critical epochs of reionization and Cosmic Dawn ($6 \lesssim z \lesssim 20$) when the  first galaxies in the Universe were forming \citep[][]{Fragos2013,Mesinger2014,Pacucci14}.  These lower-mass, accreting systems effectively \lq{}\lq{}outshine\rq{}\rq{} their SMBH counterparts, playing a possibly significant role in heating the primordial Intergalactic Medium (IGM).  However given the difficulty of directly observing X-ray emission at high-redshifts, it is local studies of NS and BH emission that gives us the best chance to characterize this emission.

The hot diffuse gas in galaxies is similarly important to understand.  The hot gaseous haloes of early-type galaxies embeds information regarding the formation and evolution of the galaxy.  The thermal structure of the ISM, in which X-ray studies play a key role, also gives key information about various physical processes affecting galaxy evolution \citep[e.g. see][]{KF2015}.  Hot gas properties are expected to differ based on e.g. mergers, ram-pressure stripping and both AGN and stellar feedback. Although there have been very detailed studies of small samples of galaxies in the local Universe, it has been impossible to do a large statistical study of the hot gas properties of early-type galaxies. 

All of our current understanding, however, is based on very small numbers of galaxies as compared to the powerhouse wide-field surveys in the optical/IR such as the Sloan Digital Sky Survey (SDSS).   We certainly know that by number, X-ray emitting normal galaxies vastly outnumber galaxies harbouring active galactic nuclei (AGN), particularly at faint X-ray fluxes.  Comparing the number counts for AGN with those for galaxies (henceforth, we refer to normal galaxies as galaxies), from the 4~Ms \Chandra~Deep Field South, \citet[]{Lehmer2012} find that galaxies become the dominant population at the faint end (S[0.5--10~keV]$<10^{-16}$ \flux; see also \citealt{Bauer2004}). 
We are on the threshold of a new generation of wide-field X-ray surveys, where galaxy samples detected in the X-ray band will final reach much larger numbers (thousands of galaxies versus 10s to 100s).   

The current state-of-the-art for X-ray surveys of normal galaxies with  \textit{Chandra} and \textit{XMM-Newton} may be broadly divided into pencil-beam (contiguous or near-contiguous fields) and serendipitous (non-contiguous fields) with the largest areas surveys being in the tens of square degrees\footnote{There has also been a nearly all-sky survey, the \XMM Slew Survey,  however this is at a fairly bright X-ray flux limit that is appropriate for studying relatively luminous and nearby AGN \citep[e.g.,][]{Dwelly2017}}.  Both the serendipitous \citep{Georgakakis2004,Georgantopoulos2005} and the pencil beam surveys \citep{Hornschemeier2003,Bauer2004,Georgakakis2006,Kim2006} have allowed the comparison of the X-ray emission to that at other wavelengths, such as infrared and optical (\ie f$_{\rm X}$/f$_{\rm opt}$), to aid in classifying AGN versus normal galaxies, and revealed important scaling relations of X-ray luminosity with global galaxy properties in normal galaxies \citep[\ie star formation rate, SFR, and stellar mass, \mstar;][]{David1992,Ranalli2003,Norman2004}. We have learned how accreting binary populations evolve over billions of years of cosmic history, demonstrating both progenitor paths for eventual gravitational wave mergers as may be detected by LIGO \citep{Maccarone} as well as showing that stellar-origin compact objects likely have significant impact on the heating of primordial IGM at $z>10$ \citep{Pacucci,Mesinger2012,Mesinger2014,GreigMesinger,Das2017}.

In this paper, we used the updated X-ray galaxy scaling relations with SFR and stellar mass, combined with newly available galaxy catalogues, to simulate X-ray observations of galaxies in a \lq{}\lq{}bottom-up\rq{}\rq{} approach to predict full-sky X-ray galaxy survey numbers. We predicted which galaxies could be detected by the  \erosita X-ray survey, which is operating on board the Spectrum-Roentgen-Gamma mission that launched on July 13, 2019 \citep{Predehl,Merloni}.

\erosita offers the next major step forward for studying galaxies at X-ray energies. \erosita will scan the entire sky eight times over four years, using an array of seven aligned mirror modules, operating in the 0.2--10~keV energy range and covering a ${\sim}1^{\circ}$ diameter field-of-view \citep[FOV;][]{Merloni}. The on-axis point-spread function (PSF) at 1.5~keV is about 15\arcsec\ (half energy width, HEW)  and degrades with off-axis angle, resulting in an average of 28\arcsec\ over the FOV. 
With a point source X-ray survey sensitivity (after four years) in the 0.5--2~keV band of $\sim1.1\times10^{-14}$ \flux\ for an average all-sky exposure\footnote{Some areas will have deeper exposure, such as the ecliptic poles; see Fig.\,\ref{fig:lxVz}} of $\sim2$ ks, \erosita will deliver the largest catalogue of galaxies detected in X-rays to date.

Approximately a decade ago, 
\citet{Prokopenko2009} made a prediction of the normal galaxy population that might be seen by \erosita surveys.  They estimated 15\,000--20\,000 ($\sim$8\,400 early-type and 7\,000--10\,000 late-type) galaxies would be detected, using a \lq{}\lq{}top-down\rq{}\rq{} technique starting from the best knowledge, at that time, of the galaxy X-ray Luminosity Function (XLF) and its evolution.  Since then, there has been great improvement in our knowledge of X-ray scaling relationships, in the quality of galaxy catalogues (permitting a \lq{}\lq{}bottom-up\rq{}\rq{} approach) and of course improvement in our knowledge of the plans for the \erosita surveys.

\begin{figure*}
\begin{center}
\includegraphics[width=1.0\textwidth]{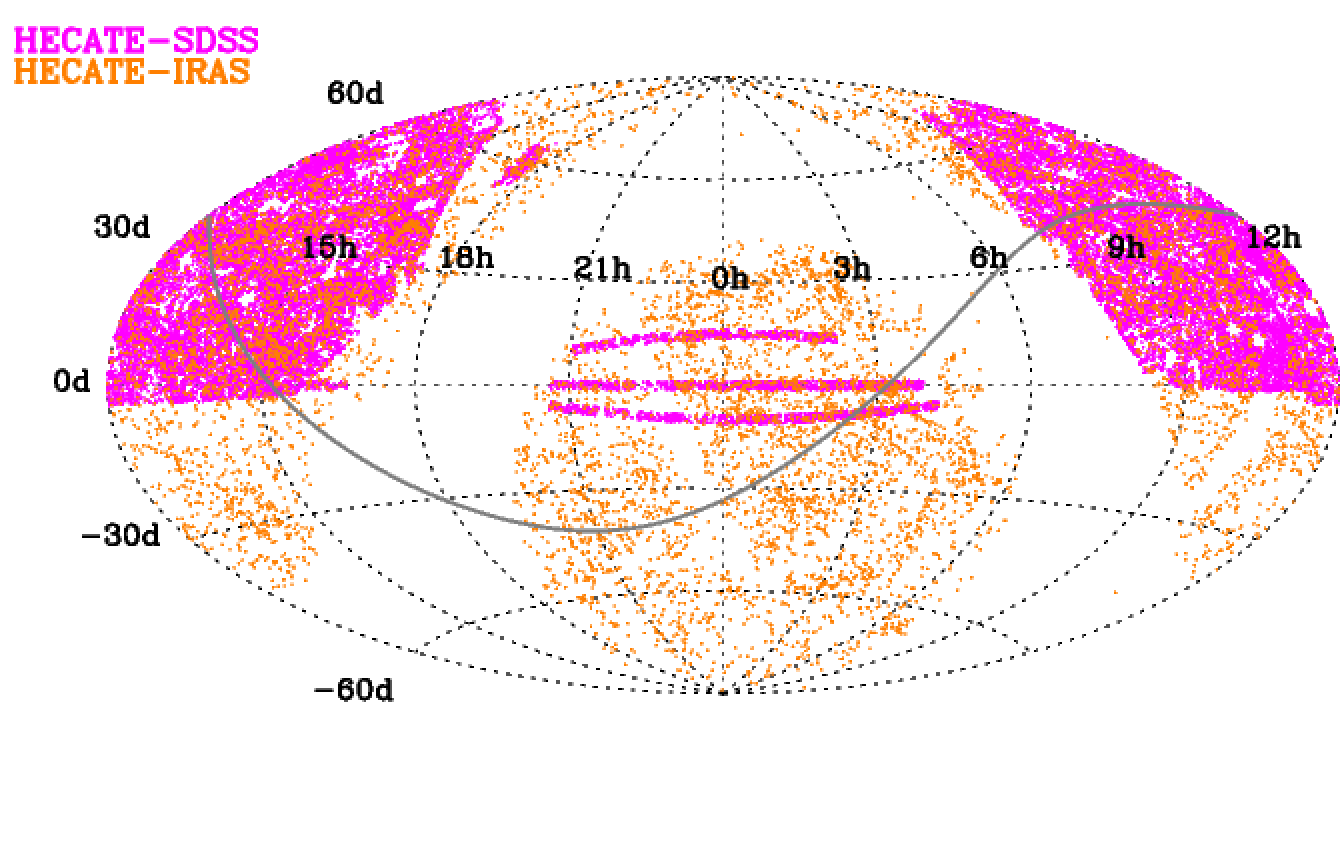}
\vspace{-20pt}
\caption{{Distribution of sources that have well-measured star formation rates (SFRs) and stellar masses as well as reliable redshifts from the HECATE SDSS (magenta) and IRAS RIFSCz (orange) catalogues. The sky map is in equatorial coordinates and centered at $\alpha=0\,\mathrm{h}$ and $\delta=0^{\circ}$, with longitudinal and latitudinal gridlines marking every 3\,h of right ascension and $30^{\circ}$ of declination. The grey line marks the separation between the \erosita-RU (above line; galactic longitude $<180^{\circ}$) and \erosita-DE (below line; galactic longitude=180--360$^{\circ}$) skies.
{\label{fig:sky}}%
}}
\end{center}
\end{figure*}

Specifically, there have been major improvements  
in our understanding of the X-ray scaling relationships for galaxies based on properties such as their SFR/stellar mass \citep[]{Mineo2012,Basu-Zych2013,Lehmer2016}.  There has also been major maturation of galaxy catalogues in the local Universe, meaning that galaxy SFRs and stellar masses are well constrained for a significant population of relatively nearby ($D\lesssim200$ Mpc) galaxies.  These two major advances mean that a bottom-up approach to modelling X-ray emission from galaxies is now possible.  In our paper we used such an approach to explore the capabilities of future X-ray survey experiments for doing normal galaxy science. 

We focussed our study on galaxies at distances 50--200 Mpc,  where we expect \erosita observations will provide significant gains over available X-ray surveys. This distance range represents the relatively unexplored volume for studying X-ray emission from normal galaxies outside the more local Universe. Our study included both early- and late-type galaxies, and used updated galaxy catalogues, revised scaling relations and state-of-the-art X-ray simulations with current \erosita response files to provide the most rigorous estimates. We describe our sample selection throughout Sect.~\ref{sec:cat}, including our methods for measuring SFRs and stellar masses in Sect.~\ref{sec:global} and screening against AGN in Sect.~\ref{sec:agncut}. In Sect.~\ref{sec:method}, we outline our techniques for simulating \erosita galaxies, using an analytic method (Sect.~\ref{sec:scale}) and also an end-to-end simulation with the Simulation of X-ray Telescopes (SIXTE) software (Sect.~\ref{sec:sixte}). We present our results and conclusions in Sects.~\ref{sec:disc} and \ref{sec:conclusion}, respectively. Throughout the paper, we assume the \citet{KroupaIMF} initial mass function (IMF) when measuring SFRs and stellar masses, and adopt $\Lambda$CDM cosmology with the following parameters: $H_0=70$ Mpc~km~s$^{-1}$, $\Omega_{\rm M}=0.3$, $\Omega_{\Lambda}=0.7$.

\section{The catalogue} \label{sec:cat}

We drew our sample from the current version of the galaxy catalogue of Kovlakas et al. (in prep), 
henceforth referred to as the Heraklion Extragalactic CATaloguE (HECATE), that contains all nearby galaxies (${\rm D} \lesssim 200\,\rm Mpc$) from the HyperLEDA database \citep{Makarov}, along with their redshifts and redshift-independent distances, where available, \forref{carefully accounting for various potential sources of errors (including infall to the Great Attractor and Shapley supercluster)}. The HECATE catalogue imposed a Virgo-infall corrected radial velocity cut of 14\,000 km/s (corresponding to $\approx$200 Mpc) on sources marked with \lq{}G\rq{}\footnote{\url{http://leda.univ-lyon1.fr/leda/param/objtype.html}}, to indicate a galaxy in the HyperLEDA catalogue. We note that the \lq{}G\rq{} object class designation not only excludes stars, star clusters, etc., but also reduces contamination from \lq{}parts of galaxies\rq{}, and \lq{}multiple galaxies\rq{}, which may appear multiply listed in other catalogues (\eg NGC 4038/NGC 4039 may appear with a variety of nomenclature: NGC 4038/NGC 4039 or NGC 4038 and NGC 4039). Therefore, restricting the sample to those with the \lq{}G\rq{} criterion provides better cross-matching with other catalogues.
The \lq{}G\rq{} criterion also allowed us to avoid regions where the X-ray emission may be dominated by the hot gas related to e.g., the intracluster or intragroup environment rather than intragalaxy X-ray emission from XRBs and/or the ISM.  Of course, X-ray emission from AGN is also a concern, and we discuss our selection against AGN in Sect.~\ref{sec:agncut} and our treatment of potential contamination in more detail in Sect.~\ref{sec:agn}.

\begin{figure}
\begin{center}
\includegraphics[width=3.4in]{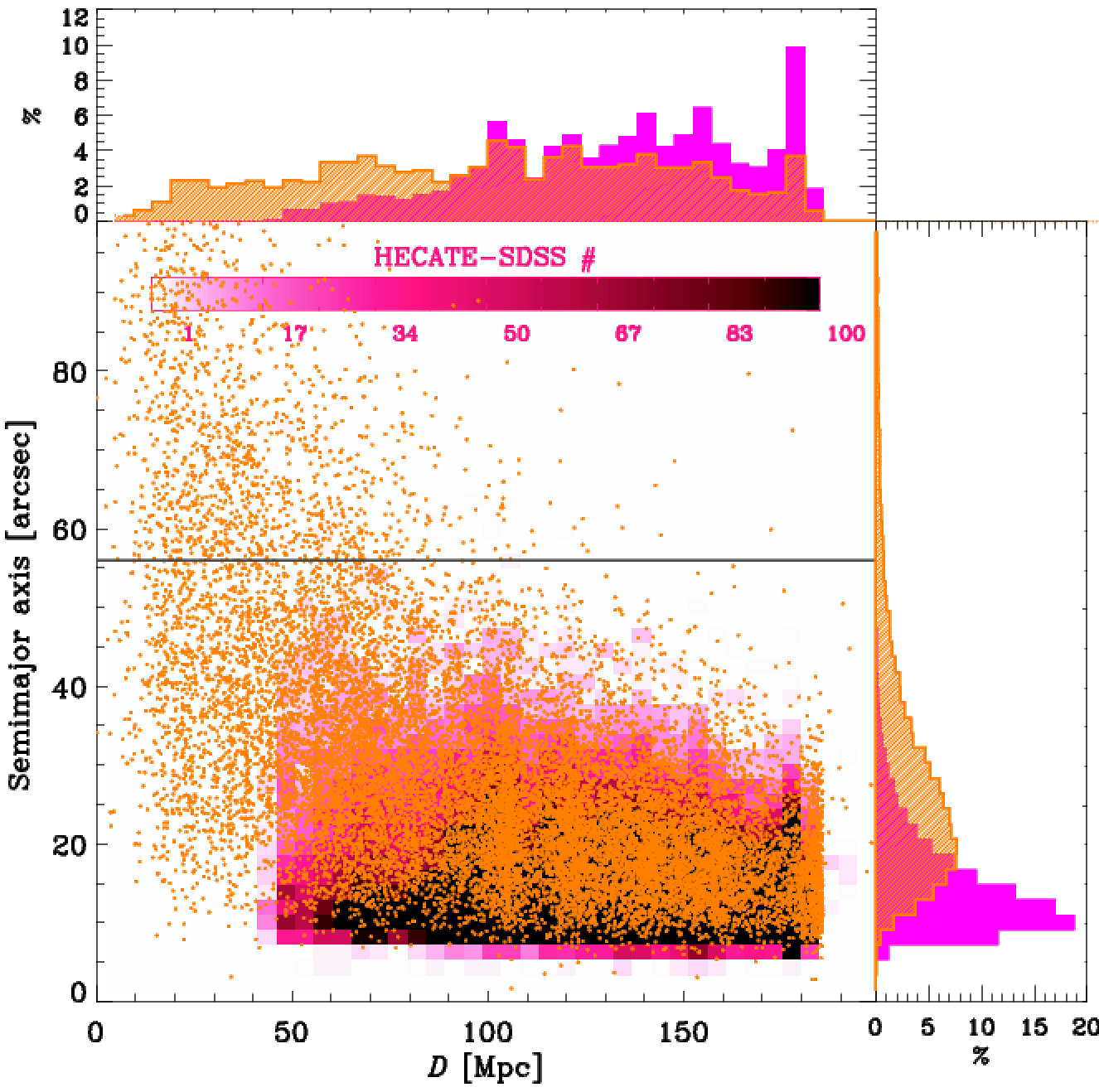}
\vspace{-0.1in}
\caption{The central plot shows the distances and semimajor axis sizes of the galaxies in the parent samples. We show the full sample of HECATE-SDSS galaxies with the magenta colour, where the colour bar describes the number scaling. The orange points mark the HECATE-IRAS sample. The histograms at right and top show the size and distance distributions, respectively, shown for the HECATE-SDSS (magenta) and HECATE-IRAS (orange) samples. The solid black line marks twice the \erosita HEW (56\arcsec). We expect galaxies smaller than this size to appear unresolved by \erosita. 
{\label{fig:d_size}}%
}
\end{center}
\end{figure}

\begin{figure}
\begin{center}
\includegraphics[width=3.5in]{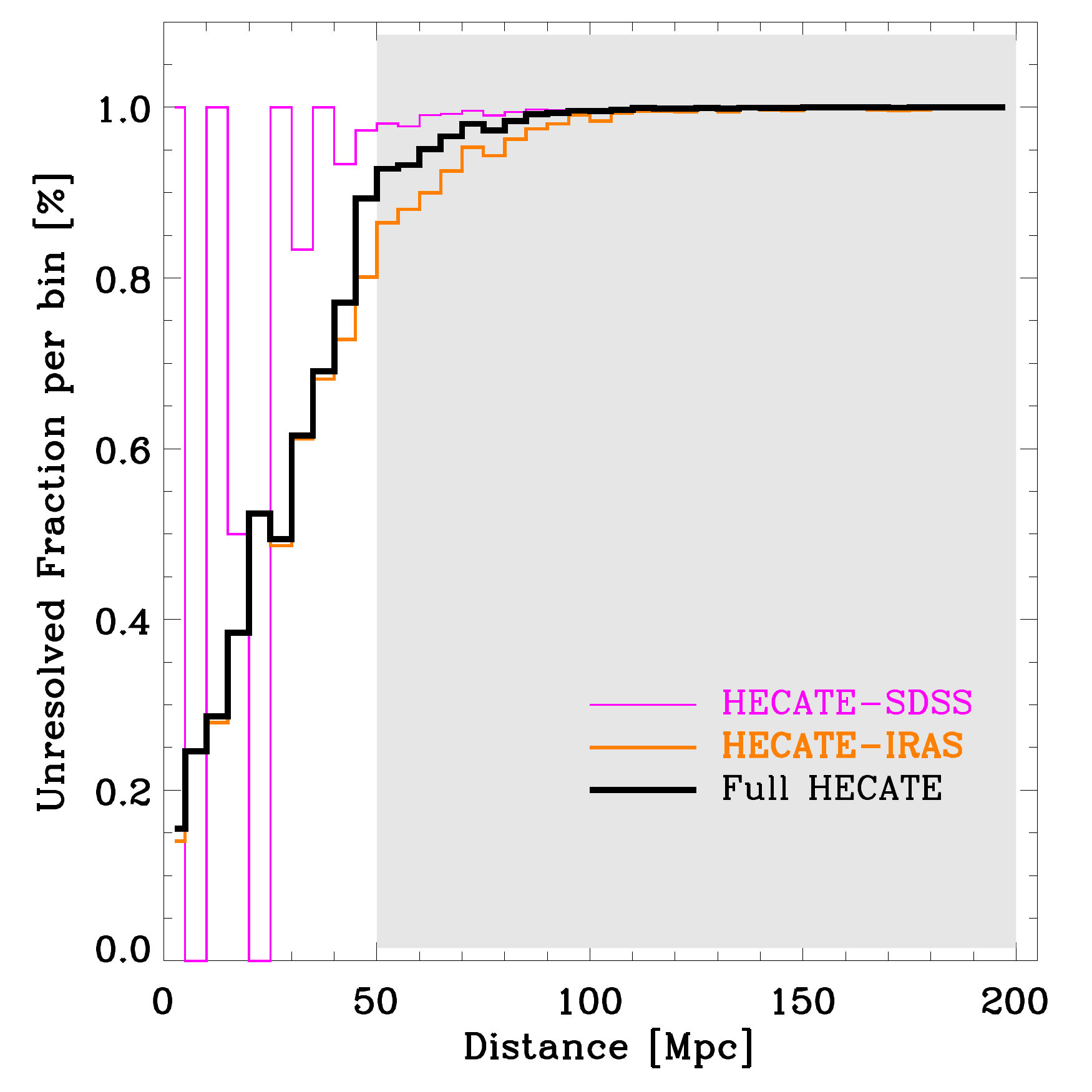}
\vspace{-0.2in}
\caption{{Fraction of galaxies that would appear unresolved by \erosita as a function of distance. In this analysis, we assume that a galaxy is unresolved by \erosita if its size is smaller than twice the \erosita HEW, \ie $<$56\arcsec. The magenta and orange lines distinguish the HECATE-SDSS and HECATE-IRAS subsamples, respectively, and the black thick line shows the total sample. We restrict our final sample to distances $>$50 Mpc where more than 90\% of the sources are unresolved.  This is also the volume that has been relatively unexplored for X-ray emission from normal galaxies in surveys.
{\label{fig:unres}}%
}}
\end{center}
\end{figure}

In addition, the HECATE galaxies were required to have coordinates with astrometric precision of $<$10\arcsec\ (note that $\sim$92\%
of the objects in the HyperLEDA catalogue have precisions of $<$1\arcsec) and were rejected if missing astrometric precisions. The method mentioned above yielded a sample of 163\,648 HECATE galaxies.

As we describe in the following two subsections, we used two subsets of the parent HECATE catalogue: the HECATE-SDSS sample (67\,875 galaxies), which includes galaxy properties from the \textit{GALEX-SDSS-WISE Legacy Catalog} \citep[GSWLC;][]{Salim2016}, and the HECATE-IRAS sample (17\,719 galaxies), where galaxy properties were measured from the \textit{Revised IRAS-FSC Redshift Catalogue} \citep[RIFSCz;][]{Wang2014}.  4\,570 galaxies are included in both subsamples (HECATE-SDSS and HECATE-IRAS)
and we discuss these further in Sect.~\ref{sec:overlap}. We applied additional cuts based on distances, 50--200 Mpc, the availability of reliable SFR and \mstar measurements, and screening against AGN, which are discussed below. \forref{In the next few subsections, we detail the sample selection steps and record the changes to the sample size with each step in Table \ref{tab:nsample}.} Our final sample contains \nfinal unique (54\,736 and 5\,476 from HECATE-SDSS and HECATE-IRAS, respectively) galaxies. Figure\,\ref{fig:sky} shows their sky distribution. 

\subsection{The HECATE-SDSS subsample}\label{sec:sdss}
The GSWLC catalogue is built primarily from spectroscopic SDSS data of the low-redshift ($z<0.3$) universe and includes galaxies that fall within the intersection of the \GALEX~+~SDSS footprints, even if they are not detected in the ultraviolet. 
The resulting catalogue is optically-selected and contains 730\,288 unique galaxies with $z=$0.01 -- 0.30 and Petrosian r-band SDSS magnitude, $r_{\rm petro}<18.0$. Imposing the HECATE-SDSS 
Virgo-infall corrected radial velocity cut of 14\,000 km/s reduces the sample by a factor of 10.  The GSWLS catalogue uses ultraviolet to optical photometry to fit the spectral energy distributions (SED), using the Bayesian approach outlined in detail in \citet[][]{Salim2016}, and measure SFRs and \mstar for the galaxies.  

\begin{table*}
\centering
\caption{Tracking changes in the sample size with each step in the analysis. The columns provide (1) row number for easy reference, (2) the analysis step, (3) the section that details the method, (4) the sample size within the HECATE-SDSS and (5) HECATE-IRAS samples, and (6) the combined sample size at the end of the step. Note that numbers within parentheses signify the total number which includes duplicate sources between the HECATE-SDSS and HECATE-IRAS samples. However, after Section 2.3 we account for these within the HECATE-SDSS sample {\em only} and therefore the numbers from this point reflect the unique numbers of galaxies. The $^{*}$ signifies those numbers which still count the duplicates.\label{tab:nsample}}
\begin{tabular}{rlcccc}
        \hline
           &  & Reference & HECATE-SDSS & HECATE-IRAS & (Total) Unique \\
         Row  & Analysis Step & Section(s) & \# & \#  & \#\\
         (1) & (2) & (3) & (4) & (5) & (6)  \\
       \hline
       \hline
       (1) & Parent sample & 2.1 \& 2.2 & 67\,875  & 17\,719$^{*}$ & (85\,594) 81\,024\\
       \hline
       (2) & $50 \leq D \leq 200$ (all) & \multirow{2}{*}{2.4} & 67\,159 & 14\,646$^{*}$ & (81\,805) \\
       (3) & $50 \leq D \leq 200$ (duplicates removed) &  & 67\,159 & 10\,186 & 77\,345 \\
       \hline
       (4) & $50 \leq D \leq 200$ $\bigcap$ SFR (Eq 1) & \multirow{ 3}{*}{2.5.1}  & 47\,831 & 5\,535 & \multirow{3}{*}{61\,067} \\
       (5) & $50 \leq D \leq 200$ $\bigcap$ SFR (``unWISE'') &  & 7\,701 & \dots & \\ 
       (6) & $50 \leq D \leq 200$ $\bigcap$ SFR (Total) &  & 55\,532 & 5\,535  & \\ 
       \hline
       (7) & $50 \leq D \leq 200$ $\bigcap$ \mstar (Eq 2) & \multirow{4}{*}{2.5.2} & 46\,608 & 3\,398 & \multirow{4}{*}{ 76\,777} \\
       (8) & $50 \leq D \leq 200$ $\bigcap$ \mstar (SED-fit) & & 20\,241 & \dots & \\ 
       (9) & $50 \leq D \leq 200$ $\bigcap$ \mstar (extrapolated $g-r$) & & \dots & 6\,530 & \\ 
       (10) & $50 \leq D \leq 200$ $\bigcap$ \mstar (Total) & & 66\,849 & 9\,928 & \\ 
       \hline
       (11) & $50 \leq D \leq 200$ $\bigcap$ \mstar $\bigcap$ SFR & 2.5.2 & 55\,312 & 5\,476 & 60\,788\\ 
       \hline
       (12) &  $50 \leq D \leq 200$ $\bigcap$ \mstar $\bigcap$ SFR $\bigcap$ AGN Screening & 2.6 & 54\,736 & 5\,476 & 60\,212 \\
        \hline
\end{tabular}
\end{table*}

\subsection{The HECATE-IRAS subsample}\label{sec:iras}
The HECATE catalogue also contains galaxies from the {\em Revised IRAS-FSC Redshift Catalogue}, which is nearly an all-sky ($\lvert b \rvert > 20 ^\circ$, \ie excluding regions near the Galactic plane) catalogue of galaxies selected at 60 $\mu$m combined with multiwavelength data from WISE All-sky Data Release \citep{WISE}, SDSS-DR 10 \citep{sdss10}, \GALEX All-Sky Survey Source Catalogue \citep{GASC}, 2MASS Redshift Survey \citep{2MRS}, and {\em Planck} Catalogue of compact sources \citep{Planck}. 
The analysis by Kovlakas et al. (in prep) 
allowed for a 10\% relative difference in the redshifts between the photometric redshifts from \citet{Wang2014} and the HECATE redshifts, and reconciled discrepant sources to create the HECATE-IRAS sample. 

The HECATE-IRAS sample complements the HECATE-SDSS sample since the former has better all-sky coverage, but tends to be biased towards brighter (i.e., closer and larger) galaxies, which is evident from Fig.\,\ref{fig:d_size} and discussed further in Sect.~\ref{sec:dcut}, where we introduced a distance criterion to restrict our study to spatially unresolved galaxies. Therefore, our analysis has missing contributions from nearby galaxies, including metal-poor dwarf galaxies, which are discussed in Sect.~\ref{sec:disc}.  

\begin{figure*}
\begin{center}
\includegraphics[width=6.2in]{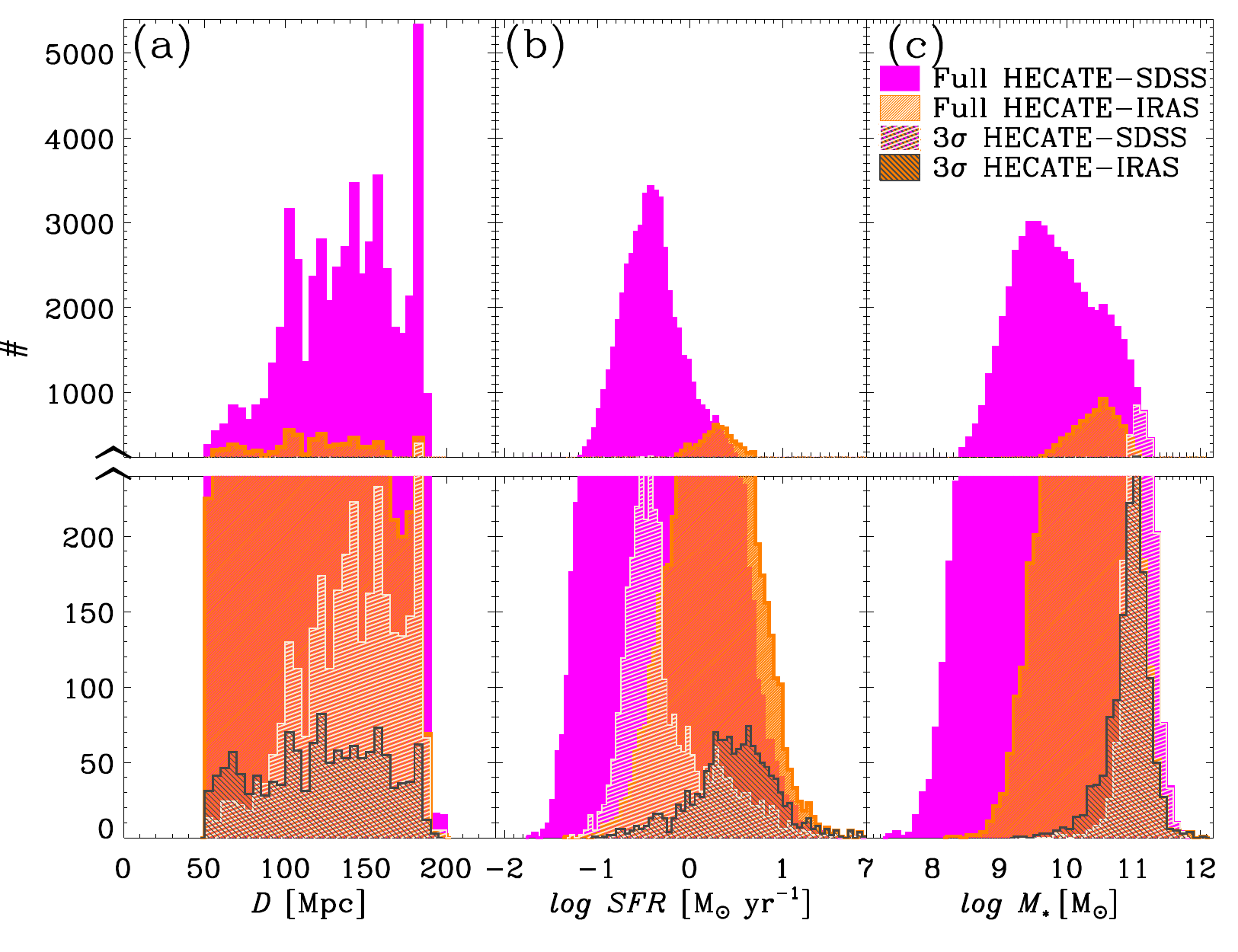}
\vspace{-0.1in}
\caption{{Histograms for distance (a), SFR (b), and stellar mass (c)
for the different samples: full HECATE-SDSS sample (magenta),~full HECATE-IRAS sample (orange),~and expected \nsig\(\sigma\) detections in the \erosita 4-year survey from the HECATE-SDSS (shaded white) and  HECATE-IRAS samples (shaded dark grey) based on the analytic simulation of pure galaxies (Sects.~\ref{sec:scale} -- \ref{sec:scatter}, i.e., no AGN components). 
{\label{fig:hist}}%
}}
\end{center}
\end{figure*}

\subsection{Subsample overlap}\label{sec:overlap}
Between the HECATE-SDSS and HECATE-IRAS samples, 4\,570 galaxies are present in both samples, 97\% of these have distances between 50--200~Mpc. These duplicates were removed from the HECATE-IRAS subsample. However, we discuss these galaxies further in Sect.~\ref{sec:mass} since they provided the opportunity for additional consistency checks. 

\subsection{Selecting spatially unresolved galaxies}\label{sec:dcut}
In this paper, we focussed on galaxies outside the immediate universe (local group and vicinity) for two major reasons. The first was to concentrate on the cosmic volume over which we expect \erosita surveys will break new ground. Pointed observations by observatories such as \XMM and \chandra have been able to conduct relatively complete surveys within 20~Mpc of relatively complete galaxy samples assembled at optical/IR/UV wavelengths. We thus expect the great gains for \erosita, as an all-sky survey, to be in the more distant universe, where it becomes impractical to target galaxies individually for follow-up. Second, for this paper we focussed on spatially unresolved galaxies, since the X-ray scaling relations that we applied (described in more detail in Sect.~\ref{sec:method}) are based upon global (integrated) emission from galaxies. In Fig.\,\ref{fig:d_size}, we show in magenta colour the distribution of distances and sizes (given by the semi-major axis in arcsec) for the HECATE-SDSS sample. The solid line marks the value for 2$\times$ the \erosita HEW \citep[56\arcsec;][]{Merloni}, below which the galaxies are assumed to appear unresolved within  \erosita. Figure\,\ref{fig:unres} shows the fraction of unresolved galaxies as a function of distance for both subsamples (HECATE-SDSS in magenta and HECATE-IRAS in orange) and the total sample (black curve). Aiming to include the largest sample of {\em unresolved} galaxies, we find that the fraction of unresolved galaxies for the final sample (black curve) flattens out between 50--200 Mpc. Therefore, we chose to restrict our sample selection to these distances. Therefore, the total sample includes 81\,805 galaxies. Of these, \noverlap galaxies overlap, resulting in 77\,345 unique galaxies with distances of 50--200\,Mpc \forref{(see rows 2 and 3 in Table \ref{tab:nsample})}. 

\subsection{Measuring galaxy global properties}\label{sec:global}
The HECATE catalogue incorporates derived galaxy properties (e.g., SFRs and \mstar) from
\citet{Salim2016} and \citet{Wang2014} for the SDSS and IRAS subsamples, respectively. However, since we strove for a consistent method of measuring SFRs and \mstar across the two subsamples, we independently measured these properties by cross-matching our sample with the \GALEX All Sky Survey \citep{GASS}, SDSS Data Release 12 \citep[DR12;][]{sdss12}, \AllWISE All Sky Survey \citep{WISEASS} and 2MASS \citep{2MASSCat} using \texttt{TOPCAT} \citep{topcat}.
We describe the SFR and \mstar measurement methods below. 
Note, however, 17\% of the galaxies do not have reliable SFRs because of missing \GALEX $NUV$ or \AllWISE $W4$ magnitudes. In addition, 568 galaxies do not have available observations to allow any method for determining \mstar (150 of these were missing SFRs as well). In order to retain the largest possible sample, we applied other methods to attain SFRs and \mstar, which are also described in more detail below.

We show the distance, SFR, and \mstar distributions for these subsamples in magenta (HECATE-SDSS) and orange (HECATE-IRAS) in Fig.\,\ref{fig:hist} panels (a) to (c). \forref{We note that, based on the completeness discussed for the HECATE-SDSS \citep[see][]{Strauss2002} and the HECATE-IRAS \citep[][]{Wang2014} samples, we expect that the samples together cover most of the galaxies in the nearby Universe ($D=50$--200~Mpc) possibly with the exception of ``missing'' galaxies pertaining to extremely faint low surface brightness, ultra-diffuse, and/or dwarf galaxies, which are unlikely to be detected by eROSITA due to their expected low SFR and \mstar. Therefore, we expect that the distribution of SFR and \mstar displayed in the middle and right panels of Fig \ref{fig:hist} effectively represent the majority of the detectable local population of galaxies. } 

\begin{figure*}
\begin{center}
\includegraphics[width=\textwidth]{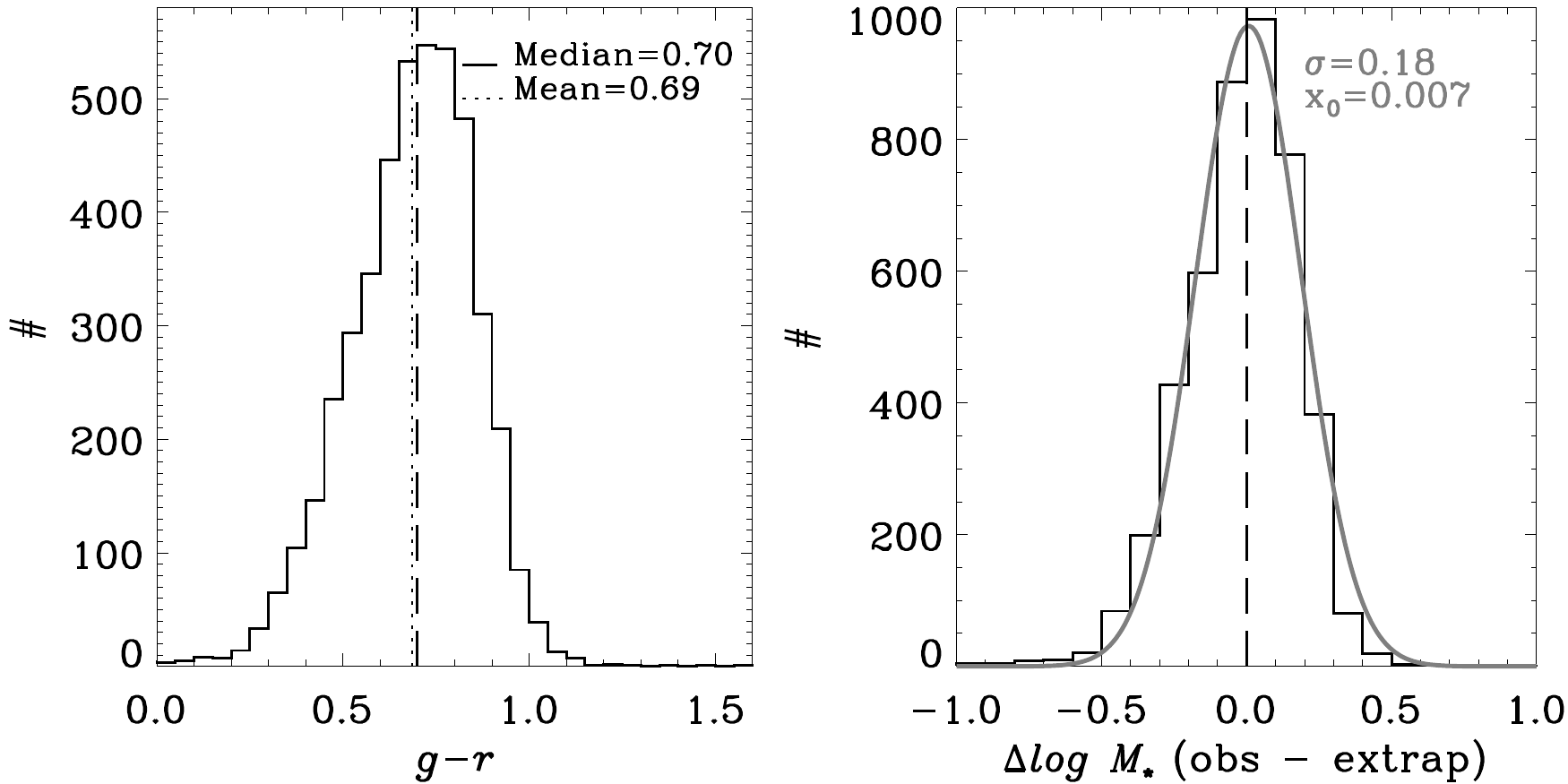}
\vspace{-0.15in}
\caption{For the galaxies in the HECATE-IRAS subsample that lack observed $g-r$ colours, we estimated the \mstar based on the \noverlap galaxies found in both HECATE-SDSS and HECATE-IRAS sample. [{\textit Left}]: Colour distribution for the overlapping sample. The mean (0.69) and median (0.70) values are very similar, and we adopted the median $g-r$ colour = 0.70 for the remaining galaxies in the HECATE-IRAS sample. In our final \mstar measurements, we used the measured colours where available (\ie for the overlapping sample) and only used the extrapolated colours for those in the HECATE-IRAS sample with missing $g-r$ colours (see discussion in Sect.~\ref{sec:overlap}). [{\textit Right}]: Comparison of the extrapolated mass measurements with those using the observed $g-r$. Grey line shows the fit to the distribution with a gaussian that is centred at $\approx$0 with $\sigma = 0.18$ dex in the $\log$\mstar values. \label{fig:extrap}}
\end{center}
\end{figure*}

\subsubsection{Star formation rates}\label{sec:sfr}
The SFRs are based on the UV$+$IR to account for stars forming both outside of and within dust-obscured regions. We applied the conversions from \cite{Bell05}:

\begin{equation}
{\rm SFR~(M_\odot~yr^{-1})}=9.8\times10^{-11}~ ({\textit L}_{IR}+3.3 {\textit L}_{UV}),   \label{eqn:sfr}
\end{equation}

where \lir~ and \luv~ correspond to the total IR (8--1000$\mu$m) and UV ($\nu L_{\nu}$ at 2800\AA) luminosities, respectively, in units of solar luminosity. SFRs were calculated using the \luv~determined from \GALEX data\footnote{We applied extinction corrections to the NUV magnitudes as given by \citet{Salim2016}.}, the \lir~determined from WISE data\footnote{Details describing the WISE All-Sky catalogue and 22$\mu$m flux measurements can be found at \url{http://wise2.ipac.caltech.edu/docs/release/allwise/expsup/index.html} and in \citet{WISE}, \citet{WISE2}, and \citet{WISEASS}.}, and Eq.\,\ref{eqn:sfr}. The factor of 3.3 is the typical correction factor that brings $\nu L_{\nu}$ (at 2800\AA) to the integrated $L_{\rm UV}$. We converted monochromatic 22$\mu$m luminosity, based on the WISE channel 4 photometry, to $L_{\rm IR}$ using the \cite{Charyelbaz} IR SED template. 

Among the $D=$~50--200~Mpc HECATE-SDSS galaxies, the method described above allows SFR measurements for $\sim$71\% of the sample. For the rest, the GSWLC catalogue provides two alternate SFR estimates (see \citealt{Salim2016} for more details): (1) based on SED-fitting, (2) based on the IR from \lq{}\lq{}unWISE\rq{}\rq{} \citep{Lang} where SDSS detections serve as forced photometry priors. 

Comparing our preferred method (Eq.\,1), applied to the majority of galaxies, with the \lq{}\lq{}unWISE\rq{}\rq{} SFRs, we find excellent agreement, with median $\Delta \log (\mathrm{SFR})=0.03$ dex [\msunyr] and $\sigma(\Delta \log (\mathrm{SFR}))=0.2$ [\msunyr{}]. Therefore, for another 7\,701 HECATE-SDSS galaxies, we adopted the SFRs given by the \lq{}\lq{}unWISE\rq{}\rq{} method in the GSWLC catalogue. SFRs are also available in the GSWLC catalogue by SED-fitting the UV to optical data. However, by comparing this method with the others described above, we found significant differences, especially at $\log \mathrm{SFR}<0$, where the other methods that included the infrared measured SFRs were 1--3 orders of magnitude higher than those from the GSWLC catalogue. Therefore, we eliminated these from our analysis.

For the HECATE-IRAS sample, nearly half of the galaxies have SFRs measured using Eq.\,1. Many of the remaining galaxies are missing SFRs due to missing UV and IR data. Therefore, our analysis includes 61\,067 
galaxies between 50--200~Mpc with valid SFRs \forref{(see rows 4--6 in Table \ref{tab:nsample})}.  

\subsubsection{Stellar masses}\label{sec:mass}
To derive the stellar masses, we used the following relation from \citet{zibetti2009}:
\begin{equation}
    \log M_\star=\log L_K + 1.119 * (g-r) - 1.257,
\end{equation} 
where $g-r$ is the colour in SDSS $g$ and $r$ bands and \mstar and $L_K$ are in solar units. Unfortunately, $L_K$ and $g-r$ colours are only available for some galaxies in our subsamples. We describe the stellar mass measurements in more detail in the next few subsections, as their treatment depended upon the subsample to which they belong. 

For the HECATE-SDSS galaxies, $g-r$ colours are available for all the galaxies, however $K$-band luminosities are only available for 69\% of the sample, therefore, the \mstar measurement either followed the methods outlined above (using Eq.\,2) or used the stellar mass derived by \citet{Salim2016}, based on SED-fits to the UV (\GALEX) and optical (SDSS) data. Comparing the masses determined by the two methods on the 69\% with $K$-band observations, we find the two methods agree very well, with average difference of $\Delta \log($\mstar$/$\msun) $<$ 0.08 dex, with no observed bias with \mstar. 

Therefore, our HECATE-SDSS subsample includes 66\,849 galaxies, \forref{with the breakdown between the two methods given in column 4 of rows 7--10 of Table \ref{tab:nsample}.} 

For the HECATE-IRAS subsample, we faced a challenge in measuring the stellar masses because $g-r$ colours are not available for 66\% of the galaxies. The majority of those with $g-r$ colours are galaxies which are in both subsamples (\ie also in HECATE-SDSS, discussed in Sect.\,\ref{sec:overlap}) and we used the average $g-r$ colours from this subsample to measure \mstar for the HECATE-IRAS galaxies that are missing observed $g-r$ colours. 

Figure\,\ref{fig:extrap} (left panel) shows the distribution of $g-r$ colour in the overlapping sample, based on the SDSS photometry. We estimated \mstar based on the median $g-r$ = 0.7 and show the distribution of ratios of the estimates compared to stellar masses measured using the observed $g-r$ colours in the right panel of Fig.\,\ref{fig:extrap}. Based on this distribution, we estimated that our method introduces uncertainties in \mstar, $\sigma \log (M_\star)\approx0.18\,\mathrm{dex}$ in the HECATE-IRAS sample, shown by the grey line. The uncertainties from this method on $\lesssim$ 10\% of the total sample is less than the scatter in the X-ray-\mstar scaling relation (given below in Eq.\,5), therefore, applying the method introduces insignificant errors on the final results. 

While the overlapping galaxies were eliminated from the HECATE-IRAS sample, they remained in the HECATE-SDSS sample and, therefore, we measured \mstar for these using their observed (true) $g-r$ values. 

\begin{figure}
\begin{center}
\includegraphics[width=3.2in]{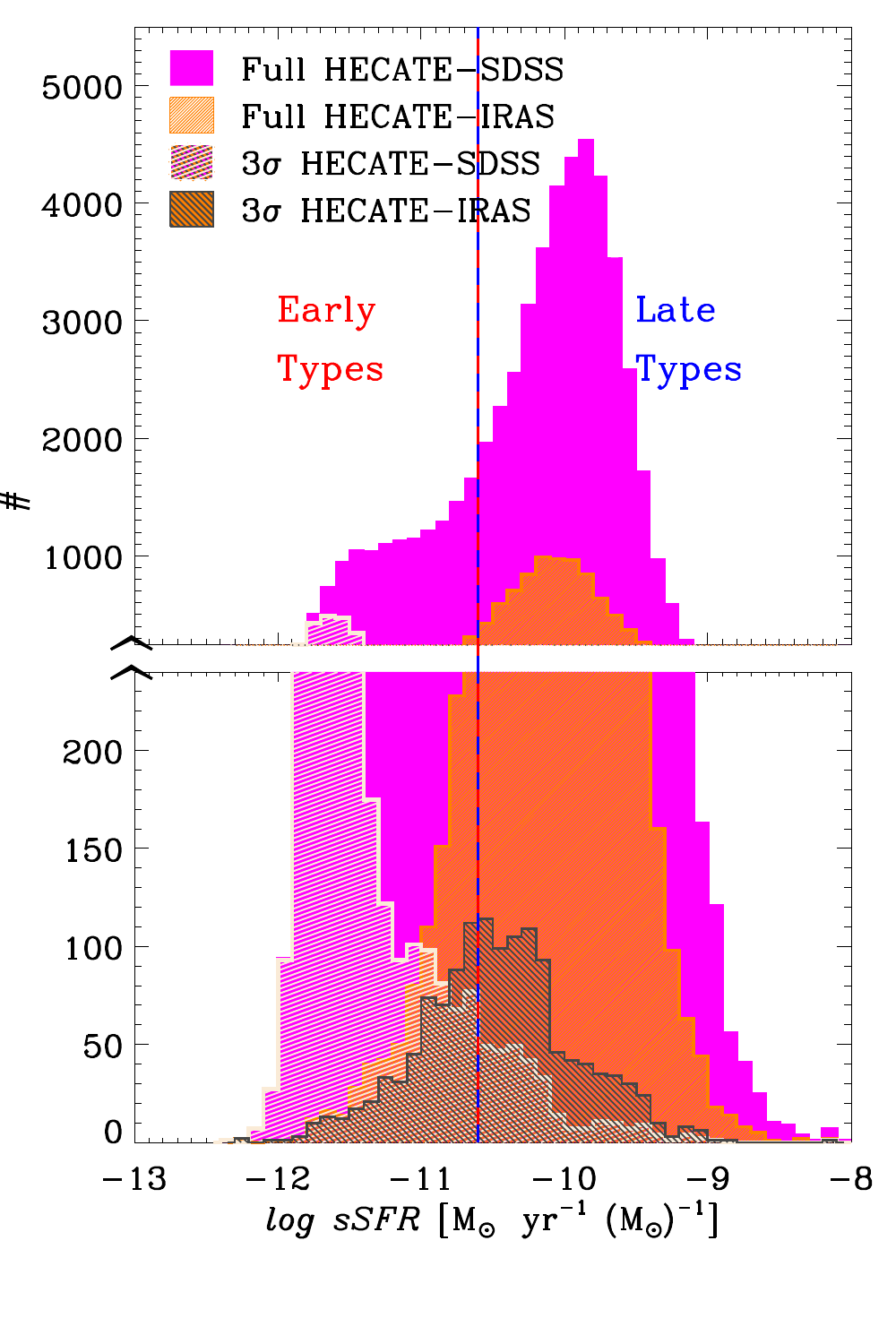}
\vspace{-0.25in}
\caption{Distribution of sSFR, with the same sample representation as in Fig.\,\ref{fig:hist}. The bimodal distribution of galaxies is apparent for the HECATE-SDSS galaxies (magenta), where $\log \mathrm{sSFR} = -10.6$ (dashed line) falls between the two peaks and separates the galaxy populations into early-types ($< -10.6$) and late-types ($> -10.6$).
{\label{fig:sSFR}}%
}
\end{center}
\end{figure}

\forref{Table \ref{tab:nsample} presents the numbers of galaxies where we measured stellar masses using either observed (row 7) or extrapolated (row 9) $g-r$ colours, or via SED-fitting (row 8). 
Of these, 79\% also have valid SFRs (row 9).}  
Any inconsistency between the different SFR and \mstar methods has no impact on the final results of this study.   

\subsection{AGN screening}\label{sec:agncut}
Since the goal of this paper is to study the X-ray emission from normal galaxies, we also want to screen against AGN. In addition, the SFR and \mstar estimates are uncertain if AGN are present, since the IR fluxes could be due to AGN emission rather than from star formation activity, \ie the assumed relation from Eq.~\ref{eqn:sfr} would not apply. 
Therefore, for the HECATE-SDSS galaxies, we screened against AGN by selecting the galaxies with \lq{}flag\_sed\rq{}$=$0 based on the SED-fitting flag from GSWLC \citep{Salim2016}, which allowed us to select galaxies that are robustly fit by galaxy spectral templates. In contrast, the catalogue identifies broad-line spectra typical of Type 1 AGN and Seyferts and those spectra with poor fits, with flags set to 1 and 2. \forref{Admittedly, screening against the broad-line AGN via this method does not remove narrow-line AGN, which may be the majority of AGN.} The HECATE-IRAS galaxies do not have any reliable AGN screening diagnostics available. However, we address potential AGN contamination in more detail in Sect.~\ref{sec:agn}. Therefore, our final sample includes \nfinal galaxies at distances of 50--200~Mpc (screened against AGN) with reliable SFRs and stellar masses. 

\section{Simulating the Normal Galaxy Population} \label{sec:method}

 For our analysis, we assumed two sources of X-ray emission within normal galaxies: the first from XRBs and the second from the hot ISM component.  Hereafter, we refer to the ISM component as the \lq{}\lq{}gas\rq{}\rq{} component.  The total X-ray luminosity for the galaxy ($L_{\rm X, gal}$) is the sum of the XRB and gas components, 
\begin{equation}
    L_{\mathrm{X, gal}}=L_{\mathrm{X},\mathrm XRB}+L_{\mathrm{X},\mathrm{gas}}
\end{equation}\label{eqn:galLx}
The applied scaling relations are sufficient for the \erosita predictions, however understanding the complexities and characterising the nature of galaxy-to-galaxy scatter in these relations are major goals for X-ray studies of normal galaxies. In particular, in-depth studies of nearby galaxies show that these scaling relations may vary with galaxy properties (\eg stellar age, metallicity). In this paper, we use the zeroth-order estimates in the scaling relations to predict \erosita detections of normal galaxies. 

\subsection{Scaling relations and spectral parameters} \label{sec:scale}
 To estimate the X-ray luminosity that originates from XRBs, we used the scaling relation from \citet[][their Eq.~15]{Lehmer2016}, where the X-ray luminosity from XRBs was further separated into the low-mass XRB (LMXB) and high-mass XRB (HMXB) components. The former scales with stellar mass (\mstar) and the latter with star formation rate (SFR), as given by 
  \begin{eqnarray}
   L_{2-10,\mathrm{XRB}}(z)&=&L_\mathrm{X}(\mathrm{LMXB})(z)+L_\mathrm{X}(\mathrm{HMXB})(z)\\
       &=&\alpha_0 (1+z)^\gamma M_\star +\beta_0(1+z)^\delta \mathrm{SFR} \label{eqn:xrb}  
 \end{eqnarray}
where $\log \alpha_0=29.30\pm0.28$, $\log \beta_0=39.40 \pm 0.08$, $\gamma=2.19\pm0.99$, $\delta=1.02\pm 0.22$. For our analysis, $z=0$ and we applied this equation using the stellar masses and SFRs discussed above to predict the X-ray luminosity due to XRBs. The XRB luminosity from the above equation is for the 2--10 keV band. 
We describe how we combine the XRB and hot gas X-ray luminosity given different bands into output 0.5--10~keV fluxes and counts, using additional spectral models, below.  

The hot gas component of galaxies differs for mass-dominated versus star-forming galaxies, with the former having hot gas scaling with the galaxy's mass, reflecting the overall size of the reservoir potential for hot ISM and the latter scaling with current star formation activity. To calculate the hot gas component, we first separated our sample into mass-dominated and star-forming galaxies, using the specific star formation rate: sSFR$=$SFR/$M_\star$. Galaxies dominated by star formation ($\log\mathrm{sSFR}>-10.1$ yr$^{-1}$) were considered star-forming and the lower sSFR galaxies were mass-dominated in our analysis\footnote{Later in the analysis, we discuss late- and early-type galaxies, divided by $\log \mathrm{sSFR}=-10.6$ yr$^{-1}$, which is a more moderate limit that better divides the galaxy types (see Fig.\,\ref{fig:sSFR})}. In Fig.\,\ref{fig:sSFR}, we show the distribution of sSFR for the two HECATE parent samples, SDSS (magenta) and IRAS (orange). For the hot gas in star-forming galaxies, we used the relation given in \citet{Mineo2012gas}, which provides the 0.5--2~keV luminosity and was converted to our chosen IMF (from Salpeter to Kroupa):
\begin{equation}
L_{0.5-2,\rm{gas}} / \rm{SFR} = (12.4 \pm 0.2) \times 10^{38} ~erg~ s^{-1} \label{eqn:SBhotgas}
\end{equation}

For mass-dominated galaxies ($\log \mathrm{sSFR} <-10.1$~yr$^{-1}$), we used the equation from \citet{KF2015}, which estimates the hot gas contribution to the 0.3--8~keV band as follows:
\begin{equation}
  \log (L_{0.3-8,\rm{gas}}/10^{40}{\rm ~erg~ s^{-1}})  = A \log (L_K/10^{11} L_{K\odot}) + B \label{eqn:hotgas}
\end{equation}
and applied their coefficients for the \lq{}\lq{}Full\rq{}\rq{} sample: $A= 2.98 \pm 0.36$ and $B= -0.25 \pm 0.11$. Since the \citet{KF2015} equation is expressed in terms of L$_{K}$, we converted this hot gas relation to stellar mass, by using the relation given by \citet{Lehmer2014},
\begin{equation}
M_\star/L_K = 0.66 M_\odot/L_{K,\odot}
\end{equation}

For the most accurate determination of detectability by \erosita, we modelled X-ray spectra as well as the total $L_{\rm X}$ of the sources. To model the X-ray spectra in these galaxies, we used a $\Gamma=1.8$ power law for the XRBs and an \texttt{APEC} model for the gas, multiplied by a \texttt{PHABS} model to account for Galactic absorption, where $N_{\rm H}$ was estimated from the sky position and the \texttt{ftools} \lq{}nh\rq{} tool. We used the default abundance model, given by \citet{angr}. 
To estimate the temperature of the gas, $kT$, from the X-ray luminosity predicted above, we applied the equation from \citet{KF2015} for their \lq{}\lq{}Full\rq{}\rq{} sample:
\begin{equation}
  \log (L_{\mathrm{X,gas}}/10^{40} {\mathrm{erg s}^{-1}}) = A \log (kT_\mathrm{gas}/0.5~{\rm keV}) + B 
\end{equation}
where $A=5.39 \pm 0.60$ and $B = 0.16\pm 0.43$.

For star-forming galaxies, \citet{Mineo2012gas} fit the spectra of hot gas in galaxies with \texttt{mekal} models with $kT=0.24-0.71$~keV, with the same abundance model. Therefore, for our sample of galaxies that have $\log \mathrm{sSFR} >-10.1$~yr$^{-1}$, we followed a similar procedure as described above, but replaced the \texttt{APEC} model with \texttt{mekal} and used $kT=0.5$ keV as a baseline value. In Sect.~\ref{sec:scatter} we uniformly sampled the $0.24-0.71$ keV $kT$ range to account for the observed range of temperature values for the hot gas component of star-forming galaxies. 

Using the distances given in the HECATE catalogue, we converted our luminosities into predicted fluxes for the galaxies. Additionally, we accounted for the encircled energy fraction of the PSF by multiplying the input fluxes by 50\% to simulate the expected flux within the PSF. These fluxes for the components in the specified bands (\ie 2--10~keV for the XRB) set the normalization for the model components. The final flux for each galaxy refers to the model flux output from the \texttt{XSPEC} model, which we output for the standard \erosita bands (0.2--0.5, 0.5--1, 1--2, 2--4.5, 4.5--10 and 0.5--10~keV). However, for simplification, henceforth our results focus on the 0.5--10~keV band, unless specified. 

\subsection{Determining \erosita detections using response files} \label{sec:resp}

We set the spectral normalization for the different spectral components (in the specified bands) using the fluxes calculated above for the gas and XRBs, as discussed in the previous section. We used the on-axis redistribution matrix file (RMF) file for all valid patterns\footnote{The RMF file, rmf01\_sdtq.fits, is available from \url{https://wiki.mpe.mpg.de/eRosita/erocalib_calibration}.} and created a combined ancillary response file (ARF) file for all 7 telescopes. The ARF file was based on the 200\,nm Al + 200\,nm PI filter configuration for telescopes 1, 2, 3, 4 and 6 and the 100\,nm Al + 200\,nm PI filter for telescopes 5 and 7\footnote{The ARF files, arf01\_100nmAl\_200nmPI\_sdtq.fits and arf01\_200nmAl\_200nmPI\_sdtq.fits, are available from 
\url{https://wiki.mpe.mpg.de/eRosita/erocalib_calibration}.}.
Using the model spectra with these files, we predicted the count rates for the 0.5--2~keV and 2--10~keV bands. We combined these counts to report 0.5--10 keV results throughout. We converted these into the expected counts for each galaxy in the \erosita all-sky survey, using the vignetting-corrected, position-dependent 4-year exposure map produced by the Simulation of X-ray Telescopes (SIXTE)  \texttt{exposure\_map} tool based on the attitude file available from the \erosita instruments installation for SIXTE. This attitude file is representative for the survey, although small deviations between this attitude file and the survey are expected due to the prolonged performance verification phase of the mission.

\begin{figure}
\begin{center}
\includegraphics[width=3.5in]{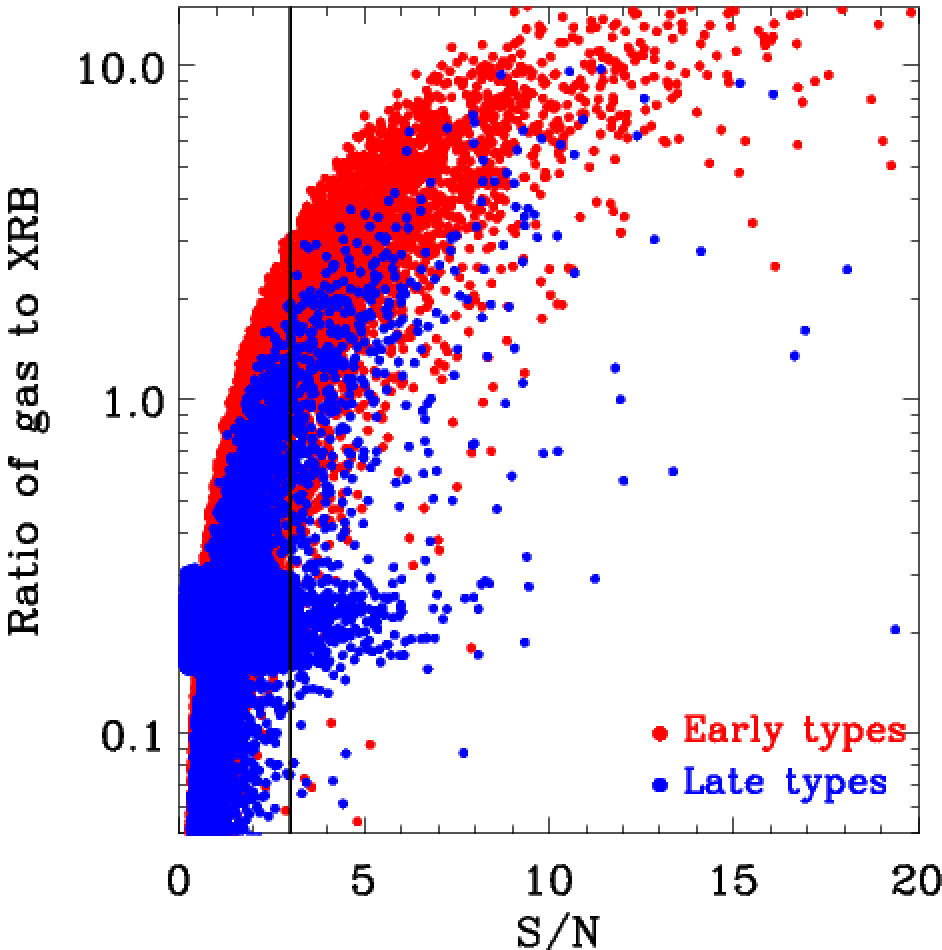}
\vspace{-0.1in}
\caption{SNR versus ratio of gas to XRB component in the expected 0.5--10~keV source counts. The vertical black line marks the 3$\sigma$ detection limit and the red and blue points mark the early- and late-type galaxies. 
{\label{fig:gas2xrb}}%
}
\end{center}
\end{figure}

To calculate the number of background counts expected in our observations, we used the model background values given for the different \erosita bands\footnote{The \erosita background model is available at \url{http://www2011.mpe.mpg.de/erosita/eROSITA_background_v8.pdf}}, given in units of counts s$^{-1}$ arcmin$^{-2}$. Therefore, to get the total number of background counts for each source, we multiplied by the same exposure used above and by the area of the galaxy. Since our galaxies are unresolved, we used the PSF area derived from the 28\arcsec~HEW for the $\sim$80\% with sizes smaller than the PSF area. For $\sim$13\,000 galaxies with larger sizes, we used the area of the galaxy, based on the available semi-minor and semi-major sizes. 

We predicted the signal to noise ratio (SNR) as $\sigma={\rm S}/\sqrt{\rm S+B}$, based on the expected source (S) and background (B) counts\footnote{As an additional test, we also calculated the Poisson S/N using Eq. 7 in \citet{LiMa83}, but found that this only made a 2\% difference in the number of $3\sigma$ detections.}. Figure \ref{fig:gas2xrb} shows the ratio of gas to XRB component in the 0.5--10~keV flux as a function of the 0.5--10~keV SNR in the two galaxy populations, \eg early types (red) versus late types (blue), with the vertical line marking the $3\sigma$ detection limit. We find that the detected galaxies are dominated by early-types, due to their higher ratio of gas to XRB emission. The cluster of late-type galaxies with gas to XRB ratios between 0.15--0.3 refer to the extreme late-type population with $\log \mathrm{sSFR} >-10.1$ yr$^{-1}$, whose prescription for the hot gas emission (see Eq.\,\ref{eqn:SBhotgas}) is different from the other galaxies (see Eq.\,\ref{eqn:hotgas}).

In the following sections, we included the effects of scatter in the scaling relations presented in this section and discuss additional caveats to \erosita \lq{}\lq{}detections\rq{}\rq{}. We show the distance, SFR, \mstar, and sSFR distributions for the detected galaxies in Figs.~\ref{fig:hist} and~\ref{fig:sSFR} for the SDSS (white/magenta striped) and IRAS (grey/orange diagonally striped) subsamples. 

\subsection{Effects of scatter on \lq{}\lq{}detectability\rq{}\rq{}}\label{sec:scatter}
Thus far, we have presented a simplified view of X-ray emission from normal galaxies, assuming that $L_{\rm X,gal}$ follows the scaling relations. In practice, galaxy-to-galaxy variations from these scaling relations can be quite large, and have been shown to vary systematically with second-order galaxy properties such as metallicity and stellar age \citep[\eg][and references therein]{Basu-Zych2013, BZ13-LBA, Douna15, Brorby2016, Lehmer2016, Lehmer2017, Lehmer2019}. In this section we explore the effect of scatter from the above scaling relations on our predictions of whether a galaxy can be detected. 

We performed Monte Carlo simulations, by creating \MCniter realizations of the XRB and gas luminosities based on the uncertainties given in the scaling relations. Specifically, each realization draws randomly from a gaussian distribution \citep[with the exception of the late type temperature, $kT$, which was uniformly drawn within the 0.24--0.71\,keV range; ][]{Mineo2012gas} that is centred on the values given in Sect.~\ref{sec:method} with $1\sigma$ standard deviations matching the given uncertainties. 

Therefore, each galaxy has a distribution of signal to noise values based on the combination of scatter within the different relations that contribute to the integrated X-ray emission from the galaxy.  

In the next section, we explore how well these predicted signal-to-noise measurements match \lq{}\lq{}detections\rq{}\rq{} within \erosita all-sky simulated data.  

\subsection{\erosita SIXTE simulations}\label{sec:sixte}
As an alternative test of the results for the \erosita 4-year survey (eRASS:8) simulations completed in the first part of Sect.~\ref{sec:method}, we produced end-to-end simulations using the SIXTE software\footnote{\url{https://www.sternwarte.uni-erlangen.de/research/sixte}} \citep{dauser08-19}. SIXTE uses an instrument model and astrophysical source models to create events files that can be used for analysis, such as image generation and source detection. It uses a Monte Carlo based approach based on simulating single photons. These are imaged onto the detector using the lab-measured energy dependent vignetting and PSF function to correctly model the imaging of each source as it passes through the field of view. While such a detailed simulation is very computational expensive, it is the most correct description of the complex \erosita instrument  in a slew survey and therefore serves as a perfect test to verify the previously obtained results.

\begin{figure}
\centering
\includegraphics[width=1.0\columnwidth]{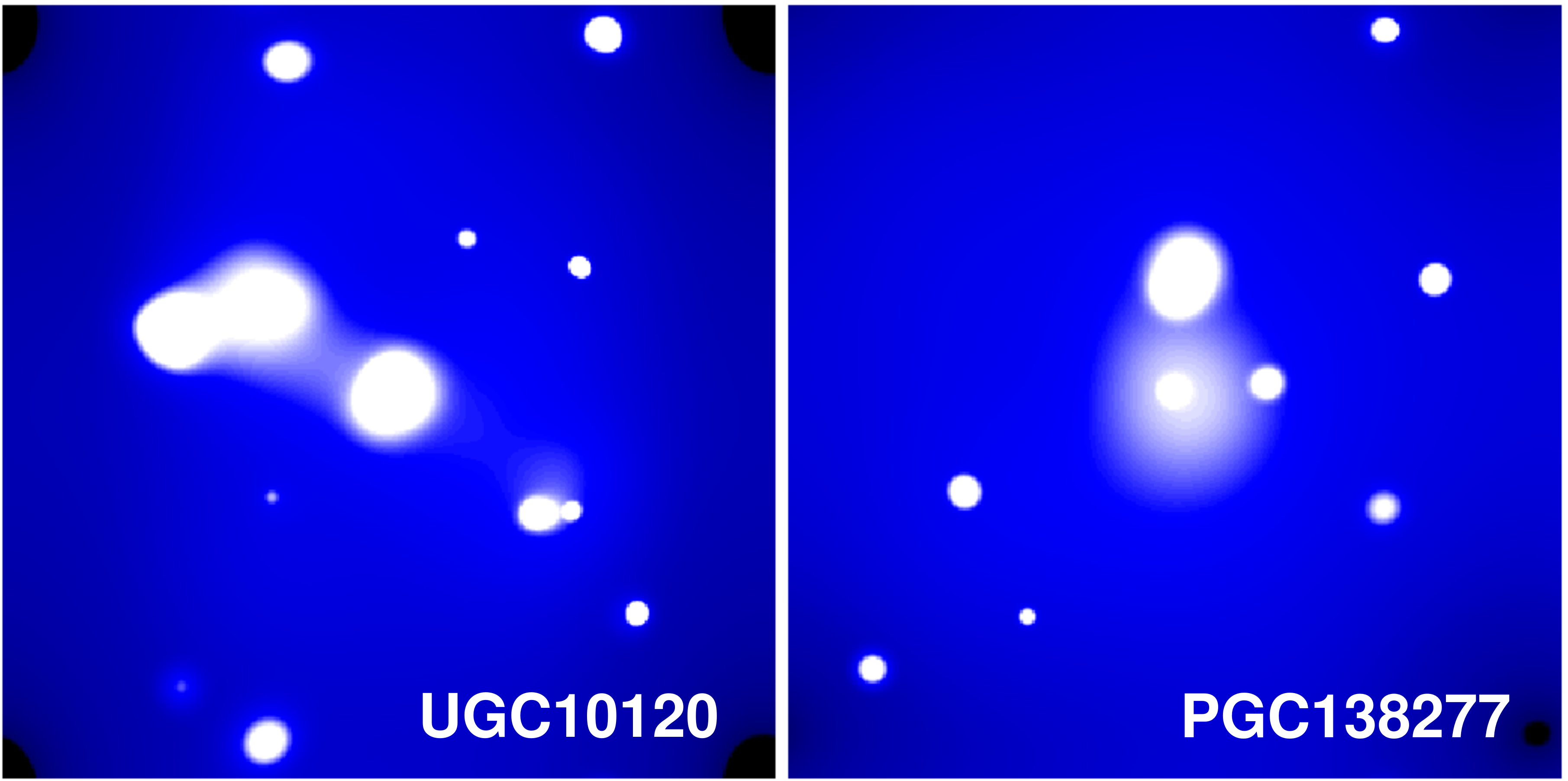}
\caption{\erosita SIXTE simulation images ($0.5-8$ keV) of galaxies detected at high (left) and low (right) significance, 54$\sigma$ and 3$\sigma$, centered on each image. Images are 1 degree on a side (\erosita FOV) and have been smoothed using \texttt{csmooth} to bring out faint features (image intensity scales differ).}
\label{fig:simimg}
\end{figure}

\begin{figure*}
\centering
\begin{tabular}{cc}
\includegraphics[width=1.0\columnwidth]{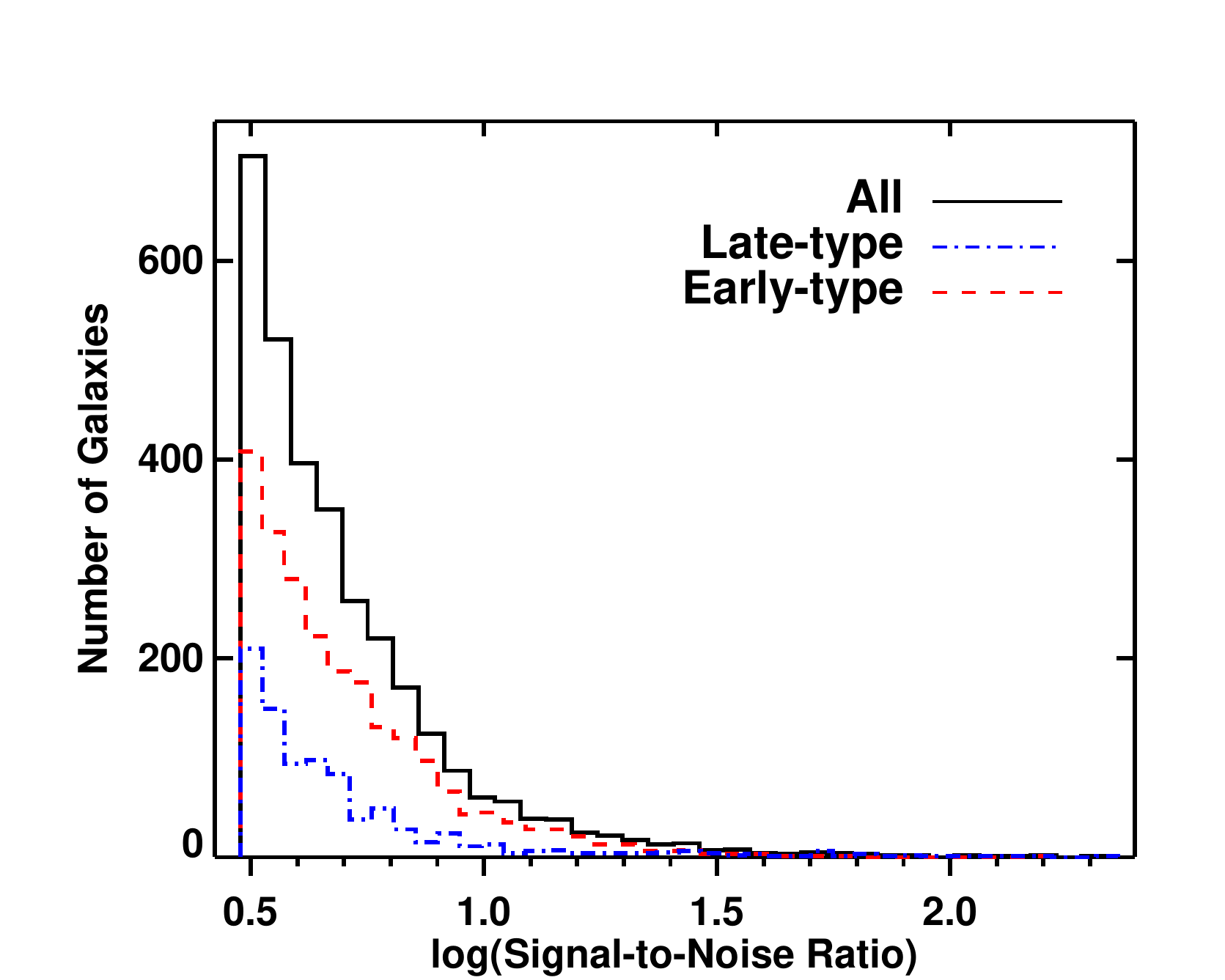}
\includegraphics[width=1.0\columnwidth]{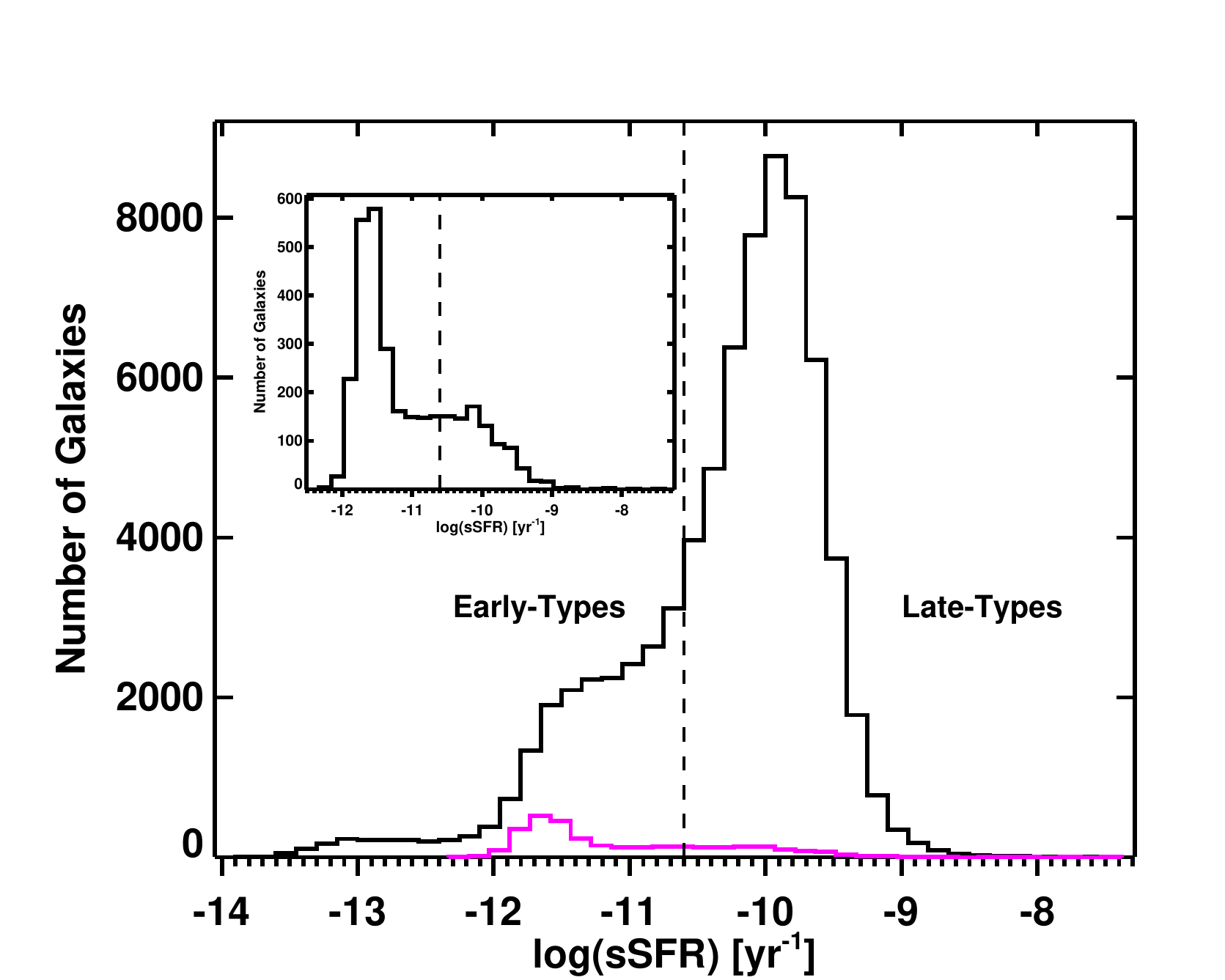}
\end{tabular}
\caption{\erosita SIXTE simulation unique galaxy detections (SNR $\geq3$, $D\geq50$ Mpc) across all energy bands. \textit{Left:} SNR for all detections (solid black line) and split by sSFR into late-type (dot-dashed blue line) and early-type (dashed red line) galaxies. \textit{Right:} Histogram of sSFR for all galaxies in the parent sample (black, Fig.\,\ref{fig:sSFR}), and all unique detections (magenta). The inset shows all unique detections (magenta histogram). The majority (67\%) of detected galaxies are early-types, based on stellar mass being the dominant driver for X-ray luminosity from the scaling relations.}
\label{fig:sixtedet}
\end{figure*}

SIXTE takes FITS-based SImulation inPUT (SIMPUT) files as inputs to the simulator, which for our purposes required the source position ($\alpha,\delta$), source flux, source spectral shape, and for various background components the source image. We followed the same procedures outlined in Sect.~\ref{sec:scale} to determine galaxy X-ray fluxes and spectra to create SIMPUT files for each of the \numin galaxies in the sample. These individual galaxy SIMPUTs served as the \lq{}source\rq{} component of the simulation. The background component was comprised of five different SIMPUTs\footnote{Provided by Thomas Dauser and Philipp Weber} that were used by \citet[][please refer to this paper for details of SIMPUT creation]{dauser08-19} to simulate the 6-month \erosita survey (eRASS:1):
\begin{enumerate}
\item Catalogue of $10^{6}$ AGN with unique source spectra \citep[and references therein]{dauser08-19, cappelluti09-07} 
\item Catalogue of $3\times10^{6}$ galaxy clusters with images and unique spectra \citep{finoguenov04-15} 
\item Catalogue of 125,000 \rosat all-sky survey point sources \citep{voges09-99}
\item The \rosat all-sky survey soft X-ray background \citep{snowden08-97}
\item Galactic ridge emission \citep{turler03-10}
\end{enumerate}

\begin{figure*}
\centering
\includegraphics[width=1.0\textwidth]{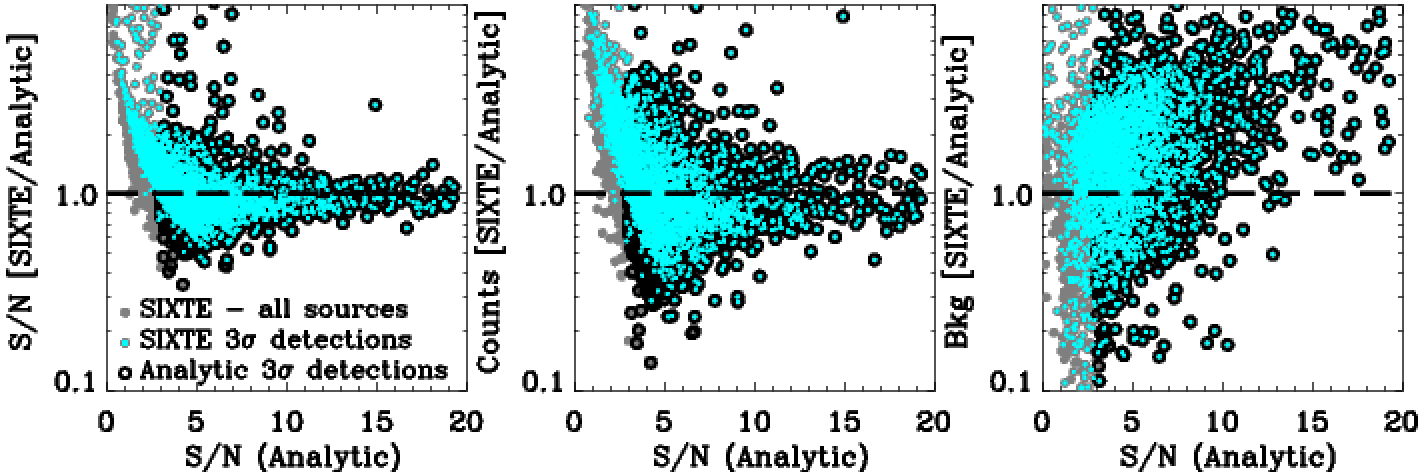}
\vspace{-0.2in}
\caption{Ratio between SIXTE simulations and the analytic approach, described in Sect.\ref{sec:method}, as a function of the analytic SNR for SNR (\textit{Left}), source counts (\textit{Middle}), and background counts (\textit{Right}). Gray points mark all sources matched to input galaxies in the SIXTE simulation, with cyan points showing those detected at $>3\sigma$. The black open circles mark the $>3\sigma$ detections in the analytic method.}
\label{fig:COMP-sixteMC}
\end{figure*}

\begin{table}
\centering
\caption{SIXTE Galaxy Detections for eRASS:8. 
Detections were restricted to distances $D\geq50$ Mpc and SNR $\geq3$.\label{tab:simres}}

\begin{tabular}{ c c c c}
\hline\hline
Full	&	Soft	&	Hard	&	All	\\
\hline
3\,069	&   3\,065	&	170	&	3\,318	\\
\hline
\end{tabular}
\end{table}
In addition, SIXTE also incorporates an estimate of the particle (non X-ray) background for \erosita, based on a flat spectrum with the expected background rate. We also included this component in our simulations. 

We simulated events for each galaxy using the input source (galaxy) and background (five separate components) SIMPUTs. To avoid long simulation times for each galaxy, we filtered the background SIMPUT files to only include background sources that were located within the \erosita FOV for a given galaxy. We used the response files noted in Sect. \ref{sec:resp} and the \erosita 4-year attitude file\footnote{eRASS\_Pc87M55\_3dobi\_att\_remeis.fits; available on the SIXTE website}\footnote{The attitude file used here is based on an earlier launch date than the real launch date, such that exposure times for individual sources will be slightly different from the real exposures. This, however, will not affect the overall statistical properties of our results.} that uses 10\,s pointing intervals that allow taking into account vignetting. For each galaxy, SIXTE created an events file for each of the 7 telescopes, which were merged using the FTOOL \texttt{ftmerge} in preparation for source detection.

We created images of each galaxy in the 0.5--10, 0.5--2, and 2--10\,keV energy bands from the output SIXTE event files using the SIXTE tool \texttt{imgev}. Source detection was completed using \texttt{wavdetect} with the scales parameter set to \lq{}\lq{}1, 2, 4, 8, 16\rq{}\rq{} for the decreased spatial resolution per pixel of 9.65\arcsec\ compared to \chandra. We selected point sources that were detected within 15\arcsec\ of a galaxy's coordinates, given that we don't expect a galaxy to be detected at a position beyond this value. In cases where more than one source was present within this radius (i.e., a background source), we selected the source with the smallest angular separation from the simulated galaxy position. We also cross-matched galaxy positions to each of the background point source catalogues within 15\arcsec, and checked to determine if a background source was mistaken for a galaxy detection. 

We selected galaxies with distances $D\geq50$ Mpc, motivated by the fact that we simulated each galaxy as a point source, and thus galaxies at this distance can be well-approximated as point sources (particularly within the average \erosita survey-mode PSF HEW of 28\arcsec\ within the FOV). Spatially resolved nearby galaxies require more detailed SIMPUT creation via inclusion of individual point sources and their spectra as well as images indicating the distribution of hot gas and the associated spectral components. This type of simulation is computationally intensive and beyond the scope of this paper, but has been completed for M31, M33, and NGC 6946 (N. Vulic). Based on Fig.\,\ref{fig:d_size}, the majority ($\sim$80\%) of galaxies have angular sizes smaller than the \erosita PSF HEW. We assume that the remaining galaxies have X-ray surface brightness profiles similar to their optical profiles, such that most of the X-ray emission is concentrated at the center of the galaxy, and thus can be represented by a point source. In Fig.\,\ref{fig:simimg} we show simulated images from SIXTE for galaxies detected at $54\sigma$ (left panel) and $3\sigma$ (right panel). 

In Table \ref{tab:simres}, we summarise the results of galaxy detections for each energy band, and the total number of unique galaxies detected across all energy bands. We calculated the SNR using the same method stated in Sect. \ref{sec:resp}. After inspecting the images we found that sources at a significance of $3\sigma$ were clearly detected and could not be attributed to background fluctuations. 
In Fig.\,\ref{fig:sixtedet} we show histograms of the SNR and sSFR for all detections ($\mathrm{SNR}\geq3$, $D\geq50$ Mpc). We also indicate the split between early-type and late-type galaxies, detecting 2\,216 and 1\,102, respectively.  

We compare the SIXTE results with those from the analytic method in Fig. \ref{fig:COMP-sixteMC}, showing the ratios of SNR (left panel), source counts (middle), and background counts (right) as a function of the analytic SNR. Overall, the agreement is fairly good. The SIXTE method yields fewer $3\sigma$ detections in the 0.5--10keV band (3\,072) than the analytic method (4\,029), which is likely due to higher background counts in the SIXTE simulations (see right panel of Fig.~\ref{fig:COMP-sixteMC}), which include additional components, \eg clusters, extended emission from the Galactic ridge along the plane.

The goal of this section is to provide a test between two different methods for detecting galaxies with \erosita. \forref{While SIXTE provides an accurate simulation of the eROSITA survey, it is also computationally intensive. The additional analyses (\eg Sections \ref{sec:scatter} and \ref{sec:agn}) would take prohibitively long with SIXTE.} Since there is overall good agreement between the results of our method, for the remainder of this paper we build on the analysis discussed in Sects.~\ref{sec:scale} -- \ref{sec:scatter} and present the results based on this analytic method. 

\subsection{AGN contamination}\label{sec:agn}
While every effort has been made to focus on the normal galaxy population, based on the optically-selected galaxy catalogue, there is some possibility that some of these sources are in fact low-luminosity or obscured AGN. In many observed cases, emission from XRBs and low luminosity AGN are both present, and while it is often unclear which dominates the X-ray emission, both contributions are significant. Therefore, our galaxy prescription described above would likely underestimate the actual X-ray emission in such sources. \forref{However, predicting the X-ray emission from accreting supermassive black holes within galaxies is not straightforward and we present this section as an illustration of how the contribution from AGN might affect the observed distribution of eROSITA-detected sources.} 

We have excluded obvious \forref{broad-line AGN} based on fits to the SED \citep{Salim2016}, as discussed previously in Sect. \ref{sec:agncut}. 
However, in this section we statistically attempt to account for the X-ray emission from potential AGN.
\begin{figure}
\begin{center}
\includegraphics[width=3.2in]{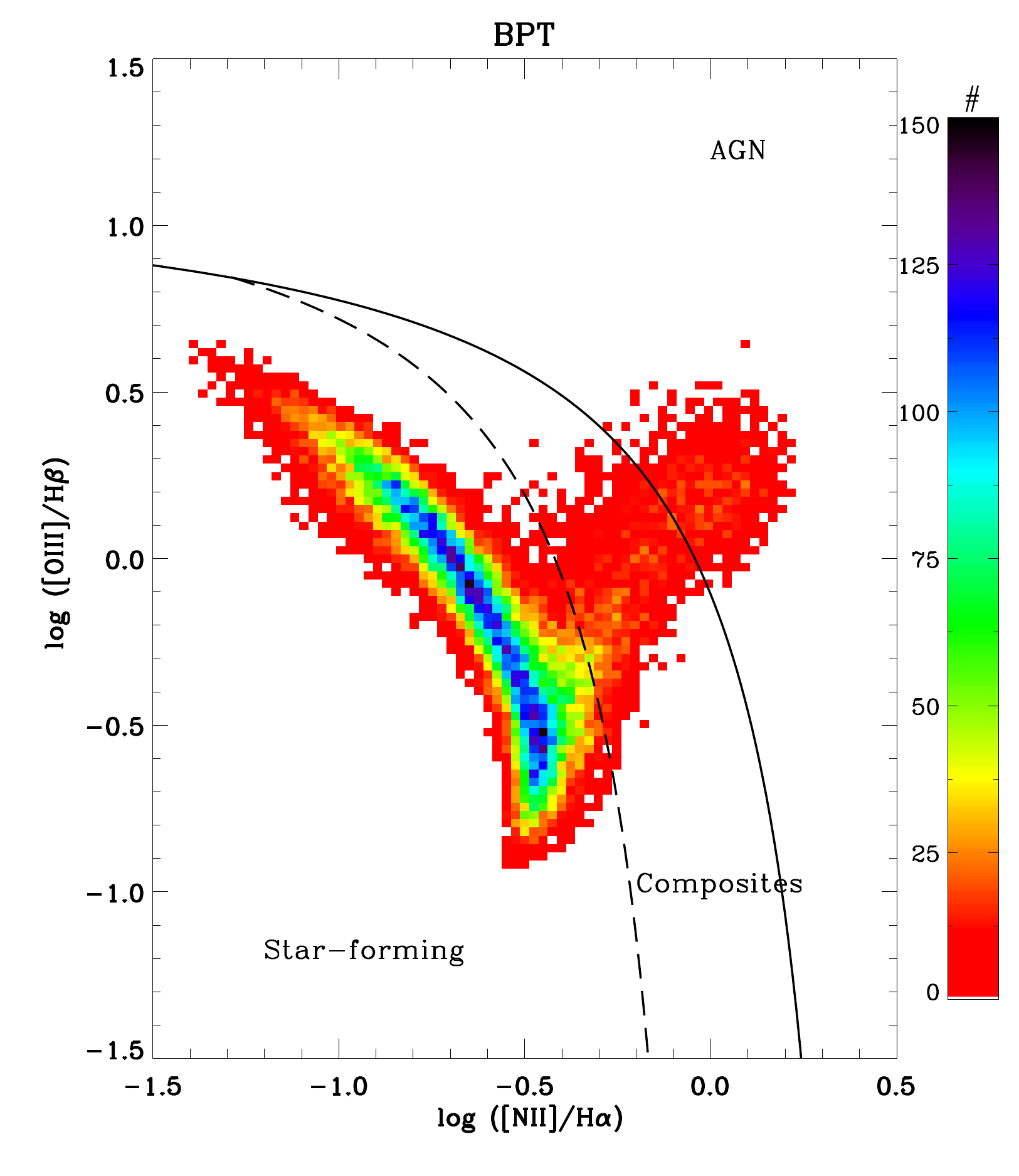}
\vspace{-0.1in}
\caption{BPT diagram, which uses the $\log [\ion{O}{iii}]/\mathrm{H}\beta$ and $\log [\ion{N}{ii}]/\mathrm{H}\alpha$ optical emission line ratios to separate the sources into star forming galaxies, composites, and AGN classes. Additionally, using $\log [\ion{S}{ii}]/\mathrm{H}\alpha$ and $\log [\ion{O}{i}]/\mathrm{H}\alpha$ line ratios help to further classify the sources into Seyferts, LINERs, and composites, which are shown in Fig.~\ref{fig:agnhist}.
{\label{fig:bpt}}%
}
\end{center}
\end{figure}
\begin{figure}
\begin{center}
\includegraphics[width=3.5in]{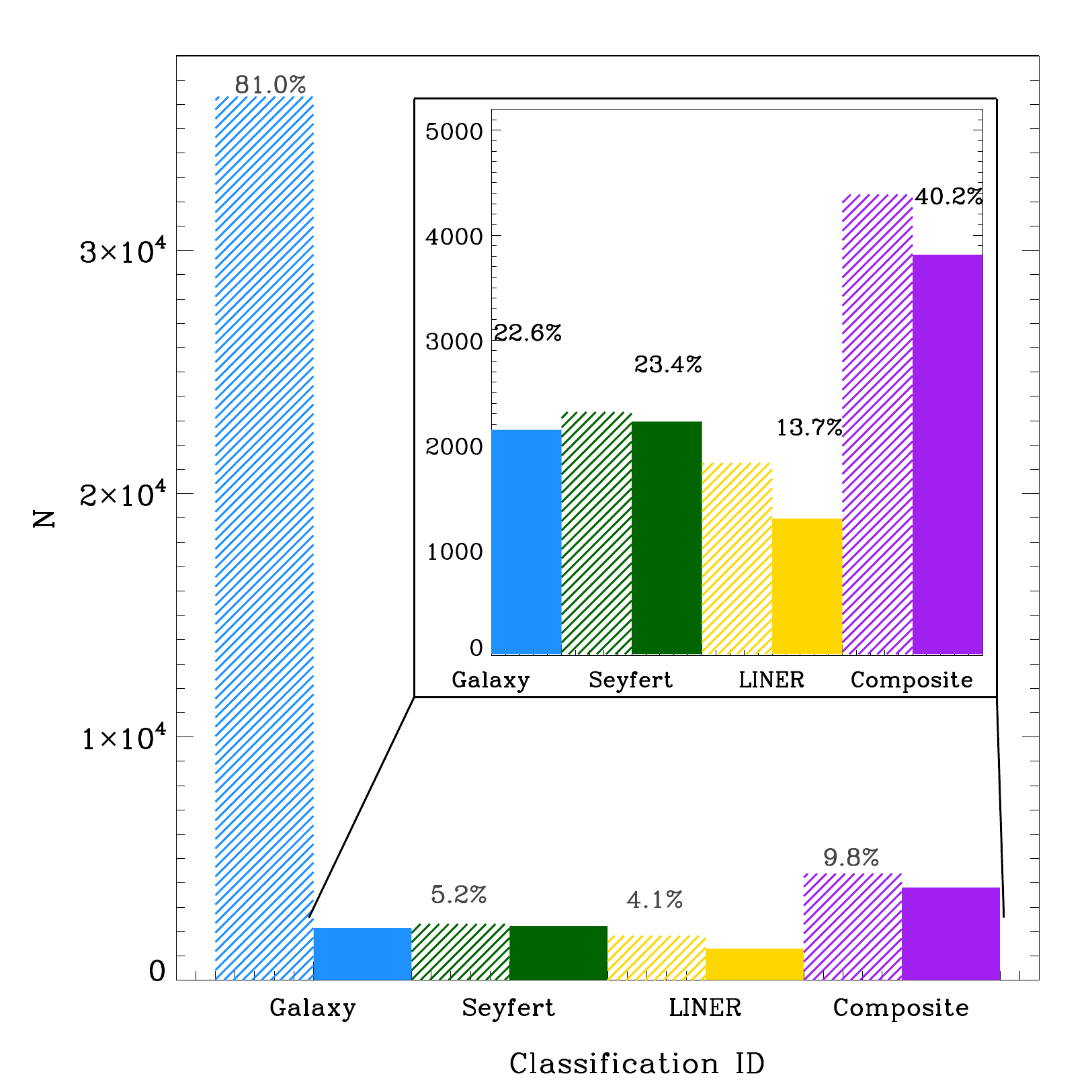}
\vspace{-0.2in}
\caption{{Histogram of galaxy classifications based on the optical emission line ratios, available for 73\% (46\,262 sources) of the HECATE-SDSS sample. For clarity, the inset provides a zoom-in to the $y$-axis for the shorter bins. The hatched and solid histograms show the input versus the detected sample. Galaxies (blue) make up the majority (81\%) of the sample, compared to 5.2, 4.1 and 9.8\% from Seyferts (green), LINERs (yellow), or composites (purple). However, galaxies only comprise 22.6\% of the detections. Significant contamination may be produced by Seyferts (23.4\%) and LINERs (13.7\%), while composite sources appear to be the dominant (40.2\%) type to be detected.
{\label{fig:agnhist}}%
}}
\end{center}
\end{figure}
The dominant contribution from star formation versus AGN processes within galaxies is often determined via optical emission line ratios \citep[][\eg BPT diagnostic]{BPT}, where star-forming galaxies are separated from AGN-dominated and \lq{}\lq{}composite\rq{}\rq{} sources by a set of theoretical and empirical curves. However, in practice, these curves are porous in the sense that X-ray detected AGN have been found within the \lq{}\lq{}purely\rq{}\rq{} star-forming region \citep[][]{AgostinoSalim, Lamassa2019}. \citet{Jones16} and \citet{Stampoulis19} present methods to extract the AGN contribution fraction based on a source's location on the BPT diagram. We note that these fractions are the contribution to the optical emission lines, not the X-ray emission. 

\forref{We also caution that the objects classified by their optical emission lines may have a complicated mix of sources producing their X-ray emission. For example, several studies have established that LINERs are a heterogeneous class \citep[see][]{Molina2018}. \citet{Eracleous2002} and \citet{Flohic2006} analyzed the \Chandra observations for several well-known, nearby LINERs and found that the X-ray emission, even within the nuclear region, are inconsistent with accreting SMBH and additional energetic sources may be responsible for powering the optical emission lines. Therefore our attempt to predict the X-ray emission from galaxies outside the \lq{}\lq{}star-forming\rq{}\rq{} locus in Fig. \ref{fig:bpt} aims to provide a rough estimate for scaling the relative contributions from XRBs and AGN.}

For our analysis, we have optical emission line data for 46\,262 galaxies from the HECATE-SDSS sample. In Fig.\,\ref{fig:bpt}, we show the BPT diagram for these galaxies, where the colour coding is scaled by the number of galaxies in each 2-dimensional bin (points) and given by the colour bar at right. The curves separating the star-forming, composite, and AGN are marked and labeled.

Using the code provided by \citet{Stampoulis19} and their definitions, we classified the galaxies using their optical emission line ratios ($\log [\ion{N}{ii}]/[\mathrm{H}\alpha]$, $\log [\ion{S}{ii}]/[\mathrm{H}\alpha]$, $\log [\ion{O}{i}]/[\mathrm{H}\alpha]$, and $\log [\ion{O}{iii}]/[\mathrm{H}\beta]$). We show the histogram of the classifications in Fig.\,\ref{fig:agnhist}: the shaded regions show all the galaxies, notwithstanding of being detected, with the percentages for each category marked in the main figure. The solid regions show the 3$\sigma$-detected galaxies with their composition stated in the inset. Based on the optical emission line diagnostics, we found that the majority of the galaxies in our detected sample (solid histograms) appear to have significant AGN (\ie Seyfert, LINER) contribution. Therefore, we estimated the total 0.5--10 keV X-ray luminosities, including potential contributions due to AGN, by performing the following Monte Carlo simulation, repeated for each galaxy \MCniter times:

\begin{table}
    \centering
    \caption{Predicted numbers for galaxies detected by \erosita. Columns (2) and (4): The number of $0.5-10$ keV detected galaxies (and percentage of \nfinal input galaxies) based on the analytic simulation of pure galaxies (Sects.~\ref{sec:scale} -- \ref{sec:scatter}, assuming no AGN contribution). Columns (3) and (5): The number of galaxies extrapolated over the full sky based on scaling Col. (2) or (4) by the area subtended by the samples (\area)  to the entire sky (41\,252 deg$^2$).}
        \begin{tabular}{|c|cc|cc|}
        \hline
         & \multicolumn{2}{c|}{eRASS:1} & \multicolumn{2}{c|}{eRASS:8}\\
         \cline{2-5}
         & simulated & full sky & simulated & full sky\\
         & \# (\%) & \#  & \# (\%) & \# \\
         (1) & (2) & (3) & (4) & (5) \\
       \hline
       \hline
        \multicolumn{5}{c}{$5\sigma$}\\
        \hline
        Total & 134 (0.2) & 491 & 1\,659 (2.8) & 6\,873 \\
        Early-Type & 115 (0.2) & 467 & 1\,412 (2.3) & 6\,321 \\
        Late-Type & 19 (0.0) & 24 & 247 (0.4) & 553 \\
       \hline
       \multicolumn{5}{c}{$3\sigma$}\\
       \hline
        Total & 467 (0.8) & 1\,924 & 4\,029 (6.7) & 16\,437 \\
        Early-Type & 406 (0.7) & 1\,796 & 3\,187 (5.3) & 14\,414 \\
        Late-Type & 61 (0.1) & 129 & 842 (1.4) & 2\,023 \\
           \hline
           \multicolumn{5}{c}{$2.5\sigma$}\\
           \hline
        Total & 743 (1.2) & 3\,051 & 5\,269 (8.8) & 21\,246 \\
        Early-Type & 647 (1.1) & 2\,847 & 3\,995 (6.6) & 18\,081 \\
        Late-Type & 96 (0.2) & 204 & 1\,274 (2.1) & 3\,165 \\
            \hline
            \end{tabular}
            \label{tab:dets}
\end{table}

\begin{enumerate}
    \item The \citet{Stampoulis19} code provides probabilities for the different classifications (\ie galaxy, Seyfert, LINER, and composite) based on the optical emission line ratios. We used these probabilities as statistical weights ($f_{\rm gal}$, $f_{\rm S}$, $f_{\rm L}$, $f_{\rm C}$ for the Seyfert, LINER and Composite classes respectively) when summing, as given in the equation below (Eq.\,\ref{eqn:total_lx}).
    \item We drew 2--10\,keV X-ray luminosities randomly from XLFs for each classification type, measured by \citet{She2017} for X-ray sources classified as Seyferts, LINERs and composites. These are designated as $L_{\rm X, S}$, $L_{\rm X, L}$ and $L_{\rm X, C}$, where the subscripts match the ones above.
    \item To maintain consistency with the methods described in Sect. \ref{sec:method}, we convert the luminosities into unabsorbed fluxes, using the galaxy's distance, and multiply this by 50\% to account for the expected flux within the PSF.
    \item We determine conversion factors from 2--10~keV to 0.5--10 keV using XSPEC with a photon index of $\Gamma=1.8$ for the AGN components, as given in \citet{She2017}, and the galaxy absorption used for Sect.~\ref{sec:method}. Therefore we determine the 0.5--10~keV model flux for each component: $F_{\rm X, S}$ (Seyfert), $F_{\rm X, L}$ (LINER), and $F_{\rm X, C}$ (composite). 
    \item The 0.5--10~keV X-ray flux is a linear combination of the separate contributions, where the galaxy contribution ($f_{\rm X, gal}$) was computed as described by Eq.\,\ref{eqn:galLx} and in Sect.~3: 
\begin{align}
F_{\rm X}= f_{\rm gal}~F_{\rm X, gal} + f_{\rm S}~F_{\rm X, S} + f_{\rm L}~F_{\rm X, L} + f_{\rm C}~F_{\rm X, C}
\label{eqn:total_lx}
\end{align}
    \item We measure the SNR from the \texttt{XSPEC} model counts and background as described in Sect.~\ref{sec:resp}.
\end{enumerate}
\forref{We note that the \citet{She2017} XLFs likely suffer from selection bias and incompleteness, and as discussed above, predicting the X-ray emission based on the optical-emission line-defined classes may be problematic.  Therefore these estimates are intended to offer a rough comparison between the scenario presented in the previous Sect. \ref{sec:global}, where 100\% of the X-ray emission is assumed to come from XRBs and the hot ISM, and this one, where we add the contribution from accreting supermassive black holes, remaining mindful of the applied assumptions and caveats.}    

\begin{figure*}
\begin{center}
\includegraphics[width=1.0\textwidth]{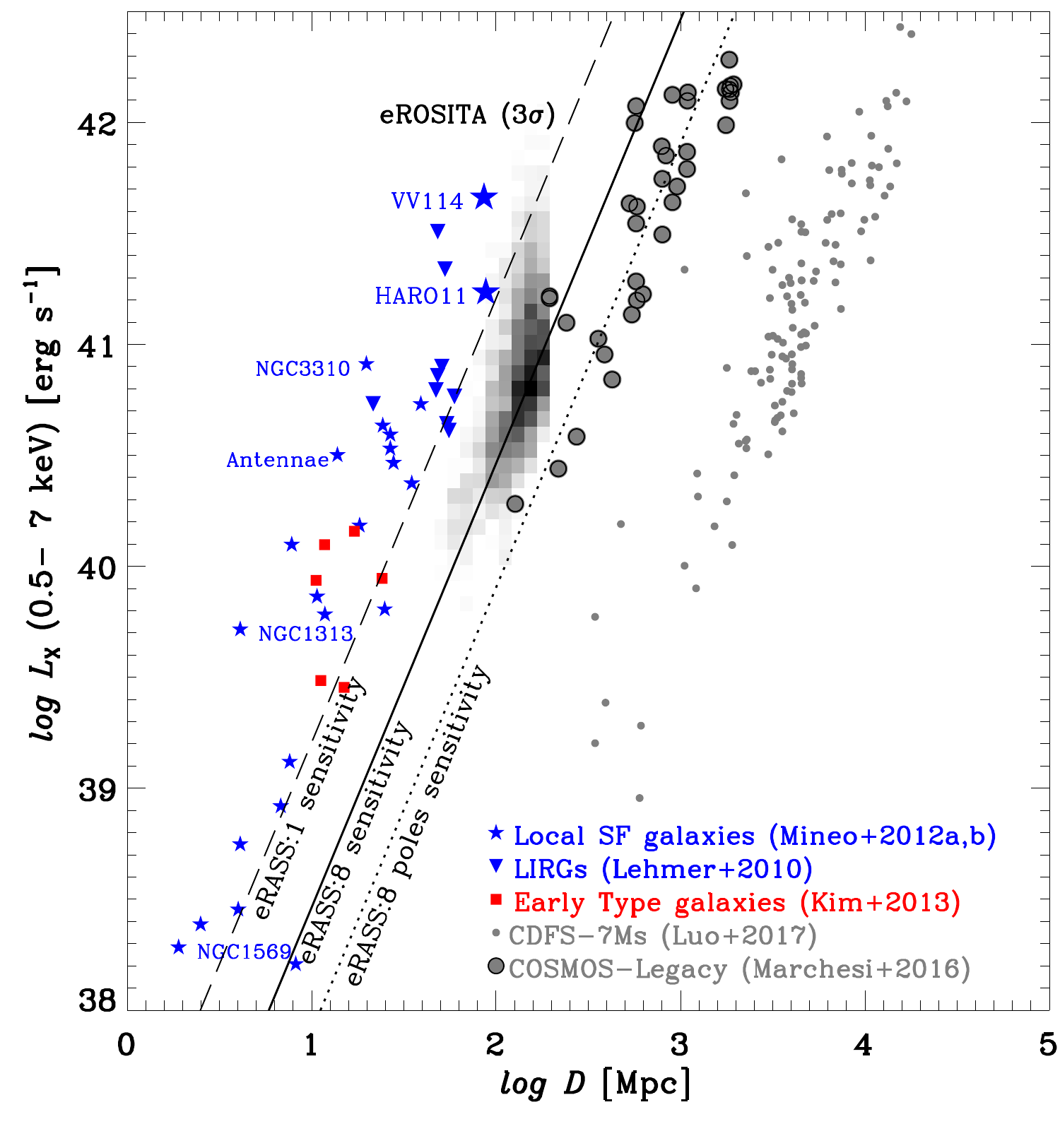}
\vspace{0.in}
\caption{{$\log$ \lx~versus distance for our \erosita simulated galaxies, shown in the background, compared to nearby star-forming \citep[blue stars; ][]{Mineo2012} and early-type \citep[red squares; ][]{Kim2013} galaxies. We mark a few well-known galaxies with their names for reference.  Luminous Infrared galaxies (LIRGs) are shown as blue triangles. At further distances, the grey points mark galaxies from the COSMOS Legacy Survey \citep[large encircled symbols;][]{Marchesi2016} and \Chandra~Deep Field - South 7~Ms survey \citep[small points;][]{Luocdfs7}. The dotted, dashed and dash-dotted lines mark the average \erosita sensitivity limits for the single all-sky survey epoch ($\approx250$ s), 4-year all-sky survey ($\approx2$ ks), and 4-year survey at the poles ($\approx20$ ks), respectively. Based on this representation, we find that nearly all of the well-studied nearby galaxies will be accessible within the first \erosita data release.  
{\label{fig:lxVz}}%
}}
\end{center}
\end{figure*}
 
\begin{figure*}
\begin{center}
\includegraphics[width=6.5in]{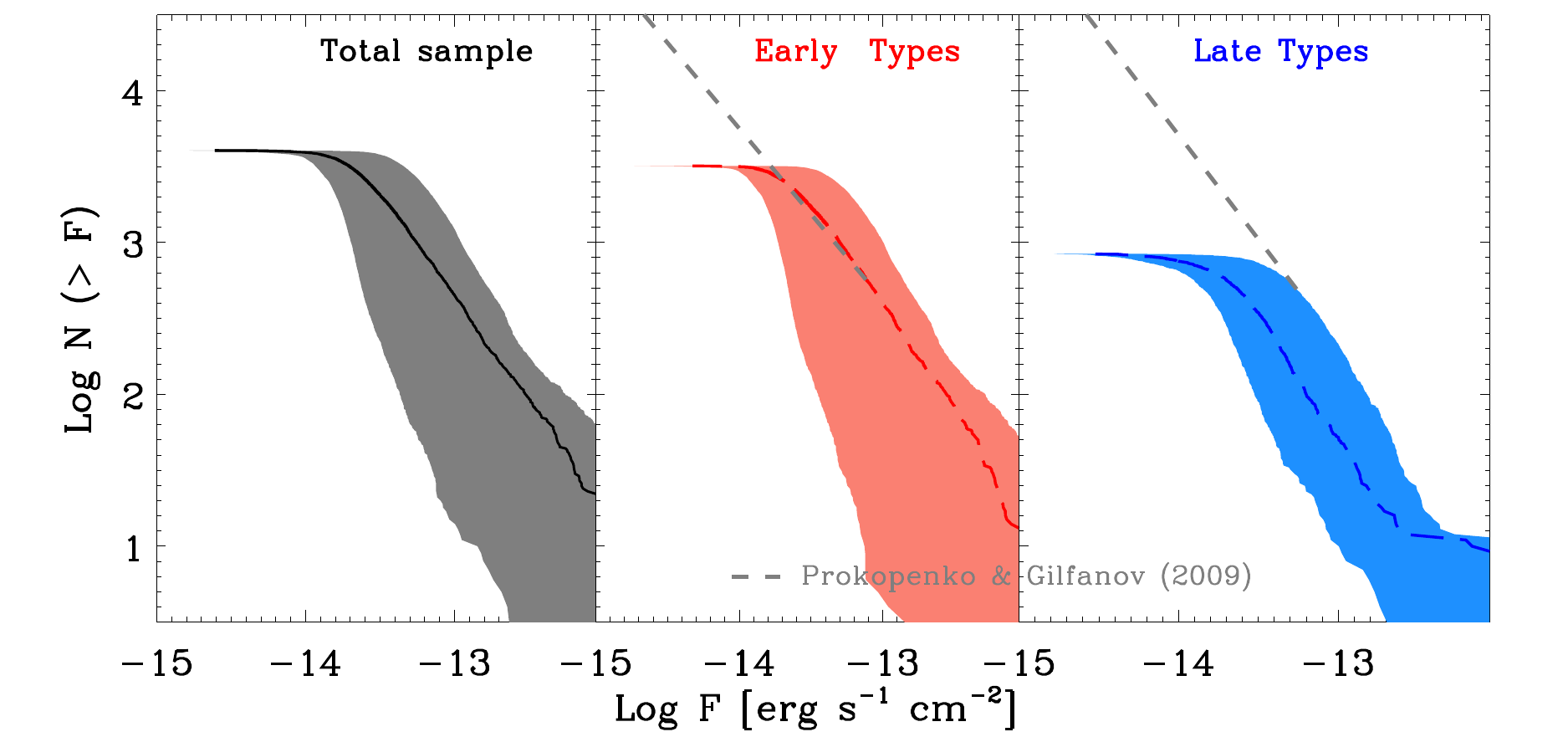}
\caption{{$\log N$-$\log S$ for all galaxies (left), early types (middle) and late types (right), showing the predicted distribution of fluxes for the \nsig$\sigma$ detections based on the analytic simulation (Sects.~\ref{sec:scale} -- \ref{sec:scatter}). The shaded regions mark the expected scatter based on the Monte Carlo simulations described in Sect.~\ref{sec:scatter}. The dashed grey line shows the \citet{Prokopenko2009} predictions for their sample of early and late types. We note that we use the sSFR to separate these two types \ie{} $\log{\rm sSFR} > -10.6$ for late-type and $<-10.6$ for early-type.
Galaxies are assumed to be pure galaxies, consisting of XRBs and hot X-ray emitting gas, with no potential contribution from AGN (in contrast to Fig.~\ref{fig:lognlogs_class}). 
{\label{fig:lognlogs}}%
}}
\end{center}
\end{figure*}

\section{Results and Discussion}\label{sec:disc}
Based on our analysis of \nfinal galaxies drawn from the HECATE-SDSS and HECATE-IRAS samples, we expect to detect \ndet galaxies\forref{, selected per methods described throughout Sect. \ref{sec:cat},} with \erosita based on the analytic simulation of pure galaxies (Sects.~\ref{sec:scale} -- \ref{sec:scatter}).  These estimates use very basic scaling relations for the estimating the X-ray emission based on SFR and stellar mass. On average, this scaling relation applies to the majority of galaxies in our sample and is sufficient for providing estimates for the predicted numbers of galaxies. Recent work has shown that the galaxy-to-galaxy scatter from these relations may be due to physical differences between the galaxies, \ie stellar ages \citep{Fragos2008, Lehmer2017,Lehmer2019} or metallicity \citep{Fragos2013, Basu-Zych2013,Douna15,Brorby2016, Lehmer2016}, yet this level of detail is beyond the scope of this paper and does not affect the overall predictions. However, our estimates indicate that the 4-year \erosita survey will detect sufficient numbers of galaxies to allow detailed investigation of X-ray emission as a function of \eg stellar age and metallicity.

Noting that the number of predictions depends on our detection (\ie signal-to-noise) threshold, we present Table \ref{tab:dets}, which offers a breakdown by galaxy type for three different SNR cuts: 5, 3 and 2.5$\sigma$. The area subtended by the HECATE-SDSS and HECATE-IRAS samples are 8\,500 and 39\,189 deg$^2$, respectively. By extrapolating our analysis \forref{of galaxies selected by the methods detailed in Sect. \ref{sec:cat}} to the entire sky by simply scaling the area to the entire sky, we estimate that \ndetfullsky galaxies may be potentially detected (\nsig$\sigma$) over the entire sky. Cols. 3 and 5 provide estimates for the different types and detection thresholds. 

\forref{Based on the SIXTE simulations (Section \ref{sec:sixte}), we expect \nsixte galaxies with SNR$\geq$3. For comparison, we extrapolate this number over the full sky to provide the lower limit estimate of \nsixtefull~galaxies.}

\begin{figure*}
\begin{center}
\includegraphics[width=6.5in]{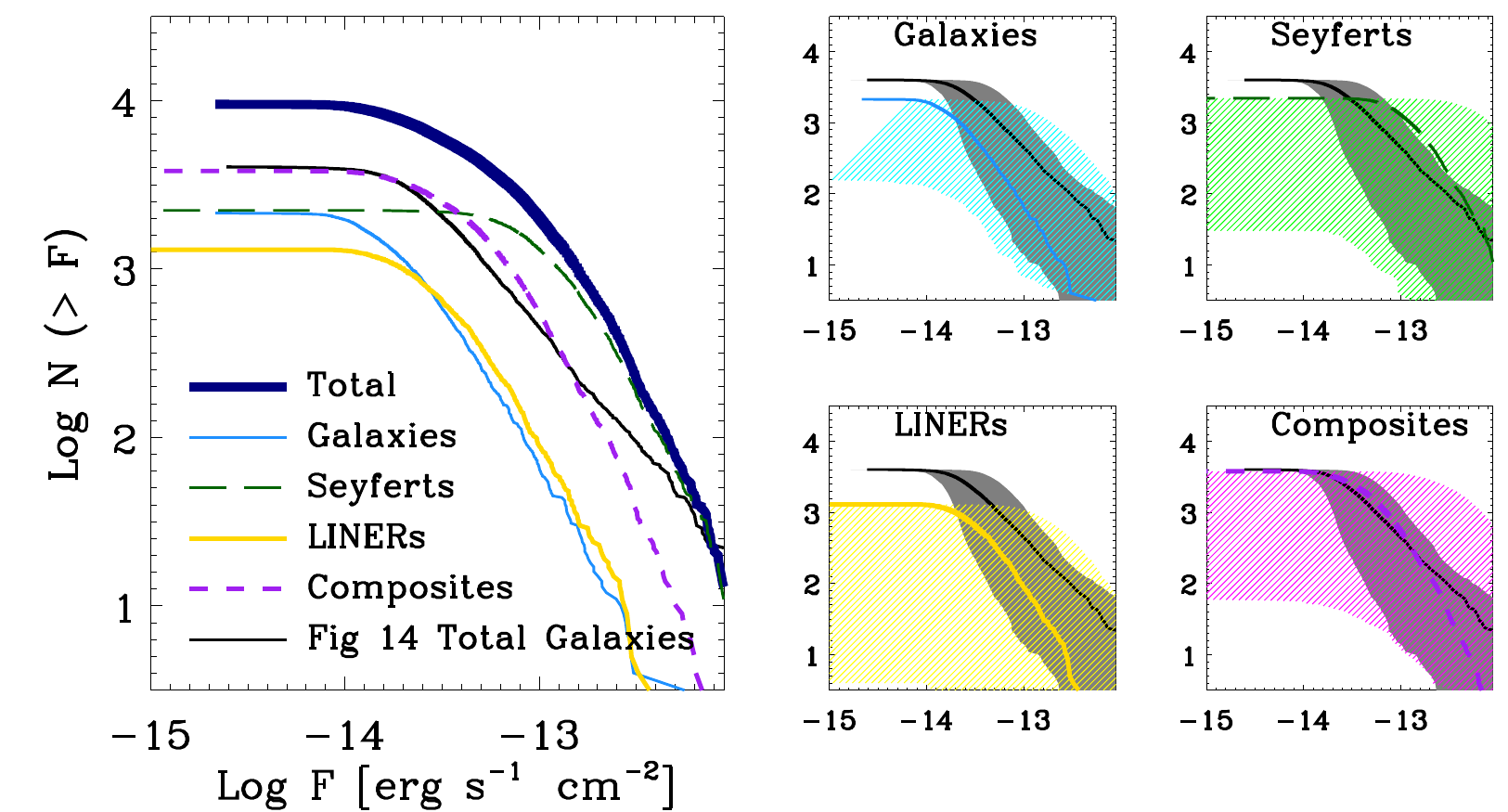}
\caption{{$\log N$-$\log S$ for the \nsig \(\sigma\)-detected sample, based on the analytic simulation (Sects.~\ref{sec:scale} -- \ref{sec:scatter}), divided into their most probable classification: galaxies (blue), Seyfert (dashed green), LINERs (thick yellow), composites (thick dashed purple). The left panel shows the summed contributions as a thick navy line to compare with the \lq{}\lq{}Fig 14 Total Galaxies\rq{}\rq{} (\ie including early and late types) curve shown in black from Fig.\,\ref{fig:lognlogs}, which assumed X-ray emission from solely hot gas and XRBs. On the right, we divide the classification types, as labelled, into separate figures to show lines and regions corresponding to the mean and 1$\sigma$ distributions based on the Monte Carlo analysis described in Sect.~\ref{sec:agn}. For all panels, the axes have identical ranges to simplify comparison and grey regions and solid black lines provide the scatter and  curves from the galaxy-only interpretation shown as \lq{}\lq{}Total sample\rq{}\rq{} in  Fig.\,\ref{fig:lognlogs}.
{\label{fig:lognlogs_class}}%
}}
\end{center}
\end{figure*}
Figure\,\ref{fig:lxVz} shows how the galaxies in the \erosita survey expect to compare in terms of distance ($D$) and luminosity (0.5--7~keV; \lx) with those from other X-ray surveys. The locus of the analytically simulated \erosita detections is shown in the background along with local star-forming galaxies \citep[blue stars;][]{Mineo2012}, including luminous infrared galaxies \citep[LIRGs, shown as blue triangles;][]{Lehmer2010}, and early-type galaxies \citep[red squares; ][]{Kim2013}. We show the distant ($D\gtrsim400$ Mpc) galaxies from the 7 Ms \Chandra~Deep Field-South \citep[CDFS;][]{Luocdfs7} as grey points. We delineate the average \erosita sensitivity limits for the first release of the all-sky survey (eRASS:1; dashed line), the full 4-year survey (eRASS:8; solid line), and that at the poles after the 4-year survey (eRASS:8 poles; dotted). In this \lx-$D$ space, \erosita galaxies occupy a poorly explored region, probing galaxies that are more distant and luminous than most of the nearby populations of galaxies.

As shown in the figure, very few galaxies appear left of the eRASS:1 (dashed) line and those that are detected must be unusually luminous (\lx $\gtrsim 4 \times 10^{40}$ \ergs) for \forref{our sample of 50--200~Mpc galaxies}. Nearby ($<$50~Mpc) galaxies, like many of the marked galaxies, were excluded from our analysis but could be potentially detected. Additionally, the first data release (eRASS:1) could include detections from incredibly rare and unique galaxies with high SFRs, like the LIRGs and low-metallicity, high-redshift analogs \citep[\eg VV114 or Haro~11, see][]{Basu-Zych2013}. 

However, the majority (85\%) of the galaxies \forref{in our analysis} require deeper exposures. Therefore, owing to its wider area coverage and sensitivity limits, we expect that the \erosita 4-year survey will be most significant in expanding our capabilities to study the X-ray emission from the complete sample of galaxies, including rare and unique populations. 

Most (74\%) of these detected galaxies are early types, based on our designation by sSFR ($< 10^{-10.6}$ \msunyr/\msun), where the X-ray emission from diffuse gas dominates the XRB component (see Fig. \ref{fig:gas2xrb}). We note that using sSFR to separate the LMXB- and HMXB-dominated galaxies has been useful for interpreting the X-ray emission from normal galaxies \citep{Mineo2012,Lehmer2016}\footnote{We note that $\log (\mathrm{sSFR}) = -10.1$~\msunyr/\msun has been used by these studies to separate LMXB- and HMXB-dominated galaxies. However, this value is fairly extreme and the HMXB-dominated galaxies correspond to very recently star-forming or starburst galaxies, rather than typical late-types. Therefore we use $-10.6$~\msunyr/\msun to separate the two types, which in Fig. 6 appears to be close to the center of the bimodal distribution in $\log(\mathrm{sSFR})$. }. However, we caution that this classification may yield different results when comparing with other methods of classification (\eg morphological or using SED-templates). 
Referring to Fig.\,\ref{fig:sSFR}, we find that detections of late-type galaxies, appearing to the right side of the dashed blue line, come mostly from the IRAS sample (orange), which has a larger fraction of nearby ($<100$ Mpc), high SFR galaxies than the SDSS sample (magenta), as panels (a) and (b)
of Fig.\,\ref{fig:hist} illustrate.
In Fig.\,\ref{fig:lognlogs}, we show the number counts expected from galaxies\forref{, based on our sample selection,}, divided into early and late types in the middle and right panels. The regions mark the 1$\sigma$ uncertainties based on the Monte Carlo realizations described in Sect.~ \ref{sec:scatter}. The uncertainties are dominated by the hot gas scaling relations in the early types (middle) and, therefore, total (left panel) galaxies. 
The prediction by \citet{Prokopenko2009}, scaled to the area of our survey, is shown in the middle and right panels as a grey dashed line for comparison. As discussed above, our galaxy type (\ie early versus late) classification is quite different from theirs, which is based on spectral templates. Their estimate is within the error region of our estimate and we expect that the main difference is driven by the different classification methods since it appears that our analysis has fewer late types compared to their predictions, but the estimate for the early types shows excellent agreement. 
As discussed in Sect.~\ref{sec:agn}, some of these galaxies may have significant contributions from AGN activity (\ie identified as Seyferts, LINERs, and composites). For the sources with measured optical emission lines, we separate $\log N$-$\log S$ into types in Fig.\,\ref{fig:lognlogs_class}. The coloured curves mark the contributions from the separate classes: \lq{}\lq{}Galaxies\rq{}\rq{} (solid cyan), Seyferts (long-dashed green), LINERs (thick yellow), and composites (purple). The thick navy line shows the sum of these contributions, which can be compared with the solid black line, which marks the \lq{}\lq{}Total sample\rq{}\rq{} curve from Fig.\,\ref{fig:lognlogs}. 

At right, each class is shown separately with the same $x$- and $y$-axes as the main plot, with regions outlining the $1\sigma$ uncertainties based on the Monte Carlo realizations described in Sect.~\ref{sec:agn}. For comparison, we show the assumption from Fig.\,\ref{fig:lognlogs}, where we assumed that all of the emission in these sources arises from galaxy processes alone (\ie no AGN contribution) as black curves (and grey regions, on the right panels).  

Based on this analysis, the composites are dominating at and above the \erosita sensitivity limit and Seyferts appear to be dominating at the brightest end of the flux distribution. \forref{We remind readers that the goal here is to provide a rough estimate of the AGN contribution, based on optical-line classifications, and caution against overinterpreting these results, given the caveats discussed in Sect. \ref{sec:agn}. However, the point here is that the pure galaxy curve (cyan) in Fig. \ref{fig:lognlogs_class} is likely a lower limit due to additional contributions from accreting supermassive black holes in some of the galaxies.} Therefore, 
the measurement of optical emission lines and careful assessment of AGN contamination will be important for \erosita studies of normal galaxies. 
 
We note a few caveats about the predictions of normal galaxies detected by the \erosita survey. Primarily, due to the conservative approach applied in this study, we expect that the numbers presented in this paper are lower limit predictions of \erosita\-detected galaxies. \forref{The selection functions are complex and largely unknown for the parent galaxy samples from which our input sample was drawn, complicating our understanding of the completeness in our study. However, these catalogs contain the largest possible compilation of nearby galaxies and incompleteness is likely only at the faintest end of the galaxy population (\eg ultra-diffuse, low surface brightness, and/or dwarf galaxies), where the predicted X-ray emission is too faint to yield eROSITA detections. Therefore if eROSITA ultimately detects significantly more galaxies than predicted, this would alter our understanding substantially of either the underlying galaxy population, as yet undetected at other wavelengths, or of the X-ray emission from faint galaxies.}

In the attempt to study a clean and well-understood sample of galaxies, we have restricted the study to unresolved galaxies by selecting a distance cut $D>50$~Mpc. Therefore, individual XRBs in Local Group and confused XRB populations at $1\,{\rm Mpc} \gtrsim D \gtrsim 50\,{\rm Mpc}$) will also be detected in the \erosita survey. For example, all the marked galaxies in Fig.\,\ref{fig:lxVz} would likely be detected and many others within our lower distance cut, at $D<50$~Mpc. Additionally, our sample excluded galaxies within groups or clusters due to the complicated diffuse hot gas contributions. Finally, we have ignored second-order corrections to the X-ray emission scaling relations that may elevate \lx/SFR due to low (sub-solar) metallicities, \eg in metal-poor starbursts.
While predicting the contribution from these galaxies is beyond the scope of this paper, these topics would lend well to the wide-survey and sensitivity capabilities of the 4-year \erosita survey. 

Our study of the normal galaxy population accessible by \erosita highlights several avenues for future galaxy science to which \erosita~will contribute, mainly by significantly increasing the numbers of detected galaxies and serving as a pathfinder for rare and bright galaxies. Data from upcoming observatories, such as The Large Synoptic Survey Telescope (LSST) and Dark Energy Survey (DES), can be combined with \erosita to offer the most complete multiwavelength survey of the Universe to date.

\section{Conclusion}\label{sec:conclusion}
The estimates, presented in this paper, are based on the zeroth-order scaling relations. However, a major goal of X-ray studies of normal galaxies is to characterise the nature of the galaxy-to-galaxy scatter observed in these relations. \erosita, by boosting the number of X-ray detected galaxies current by a factor of $\sim 100$, will offer the statistical sample required to systematically study the effects of X-ray emission on other galaxy properties such as metallicity and stellar ages.

Based on its expected 4-year survey coverage and sensitivity, \erosita~is expected to significantly increase the number of detected galaxies and further our understanding of the X-ray emission produced in galaxies. In this study, we start from available multiwavelength catalogues, HECATE-SDSS and HECATE-IRAS, to select galaxies between 50--200~Mpc. We focus on galaxies at these distances to apply scaling relations that estimate the X-ray emission based on galaxy-wide SFRs and stellar masses, and because at these distances \erosita will likely have the most relevant impact on X-ray galaxy studies. Based on this analysis, we summarise our main results as follows: 
\begin{itemize}
\item Of the \nfinal~galaxies in our final sample\forref{, selected as described in Sect. \ref{sec:cat},}, \erosita will detect \forref{\nsixte ($\sim$5\%)} galaxies at $>3\sigma$ significance. Extrapolating over the full sky, we predict that \erosita will detect \forref{\nsixtefull~galaxies}, which will be divided equally between the \erosita-DE and \erosita-RU data.
\item Most ($\sim70$\%) of these detections will be galaxies with $\log {\rm sSFR} < -10.6$, 
which we designate as early-type galaxies for our study.
\item Due to our conservative approach, our estimate is a lower limit for the following reasons:
\begin{enumerate}
    \item \erosita is also likely to detect nearby ($<50$~Mpc) galaxies or their spatially-resolved components (\ie XRBs). 
    \item We have excluded galaxies residing in or near groups or clusters from our analysis due to the complicated hot gas contribution from these environments.
    \item We did not apply the metallicity-dependent scaling of X-ray emission with SFR, which could lead to higher X-ray luminosities and, therefore, lead to potentially more detections of low-metallicity (and late-type) galaxies. 
\end{enumerate} 
\item As shown in Fig.\,\ref{fig:lxVz}, the \erosita 4-year survey probes a unique region in terms of distance and X-ray luminosity compared to other galaxy studies. While the eRASS:1 data (\ie first data release including 6 months of survey data) may discover extremely rare, X-ray luminous galaxies, the full 4-year survey will ultimately offer the most complete and statistically robust X-ray study of galaxies. 
\item Given the expected number counts (see Fig.\,\ref{fig:lognlogs_class}), multiwavelength follow-up observations, especially optical spectroscopy, will be required to help distinguish normal galaxies from other classes of galaxies (\eg Seyferts, LINERs, and composites). 
\end{itemize}

\section*{Data Availability}
Most of the data underlying this article is available via publicly available catalogs, \eg HyperLEDA (\url{http://leda.univ-lyon1.fr/leda/param/objtype.html}), the GALEX-SDSS-WISE Legacy Catalog \citep[\url{https://salims.pages.iu.edu/gswlc/}; see][]{Salim2016} and the Revised IRAS-FSC Redshift Catalogue \citep[\url{http://www.astro.dur.ac.uk/\~lwang83/RIFSCz.tar.gz}; see][]{Wang2014}. Any additional data request based on this article can be requested by contacting the corresponding author.

\section*{Acknowledgements}
We thank the referee for helpful suggestions that improved the clarity of this manuscript. We gratefully acknowledge Violet Replicon for her contributions to this work, Mackenzie Jones for sharing code and data on AGN fractions, Thomas Reiprich for providing useful suggestions that significantly improved the analysis, and Philipp Weber for useful discussions and provision of some of the SIMPUT catalogues provided here. We acknowledge funding support from NASA GSFC through its Internal Research And Development (IRAD) program.
This work was partially funded by the Bundesministerium f\"ur Wirtschaft und Technologie based on a resolution of the German Parliament through Deutsches Zentrum f\"ur Luft- und Raumfahrt grant 50\,QR\,1603.
We acknowledge use of public data from the Sloan Digital Sky Survey. Funding for SDSS-III has been provided by the Alfred P. Sloan Foundation, the Participating Institutions, the National Science Foundation, and the U.S. Department of Energy Office of Science. The SDSS-III website is \url{http://www.sdss3.org/}. This publication makes use of data products from the Wide-field Infrared Survey Explorer, which is a joint project of the University of California, Los Angeles, and the Jet Propulsion Laboratory/California Institute of Technology, funded by the National Aeronautics and Space Administration. This publication makes use of data products from the Two Micron All Sky Survey, which is a joint project of the University of Massachusetts and the Infrared Processing and Analysis Center/California Institute of Technology, funded by the National Aeronautics and Space Administration and the National Science Foundation. This research has made use of the NASA/IPAC Extragalactic
Database (NED) which is operated by the Jet Propulsion Laboratory, California Institute of Technology, under contract with the National
Aeronautics and Space Administration.



\bibliographystyle{mnras}





\bsp	
\label{lastpage}
\end{document}